# A Fast Statistical Method for Multilevel Thresholding in Wavelet Domain

Madhur Srivastava [a], Prateek Katiyar [a1], Yashwant Yashu [a2], Satish K. Singh[a3], Prasanta K. Panigrahi [b*]

[a] *Jaypee University of Engineering & Technology, Raghogarh, Guna – 473226, Madhya Pradesh, India*

[b] *Indian Institute of Science Education and Research- Kolkata, Mohanpur Campus, Mohanpur - 741252, West Bengal, India*

---

**ABSTRACT**

An algorithm is proposed for the segmentation of image into multiple levels using mean and standard deviation in the wavelet domain. The procedure provides for variable size segmentation with bigger block size around the mean, and having smaller blocks at the ends of histogram plot of each horizontal, vertical and diagonal components, while for the approximation component it provides for finer block size around the mean, and larger blocks at the ends of histogram plot coefficients. It is found that the proposed algorithm has significantly less time complexity, achieves superior PSNR and Structural Similarity Measurement Index as compared to similar space domain algorithms[1]. In the process it highlights finer image structures not perceptible in the original image. It is worth emphasizing that after the segmentation only 16 (at threshold level 3) wavelet coefficients captures the significant variation of image.

*Keywords:* Discrete Wavelet Transform; Image Segmentation; Multilevel Thresholding; Histogram; Mean and Standard Deviation; .

*Corresponding author, Mob. : +91 9748918201.
E-mail addresses: pprasanta@iiserkol.ac.in, (P.K. Panigrahi).

---

## 1. INTRODUCTION

Image segmentation is the process of separating the processed or unprocessed data into segments so that members of each segment share some common characteristics and macroscopically segments are different from each other. It is instrumental in reducing the size of the image keeping its quality maintained since most of the images contain redundant informations, which can be effectively unglued from the image. The purpose of segmentation is to distinguish a range of pixels having nearby values. This can be exploited to reduce the storage space, increase the processing speed and simplify the manipulation. Segmentation can also be used for object separation. It may be useful in extracting information from images, which are imperceptible to human eye [2].

Thresholding is the key process for image segmentation. As thresholded images have many advantages over the normal ones, it has gained popularity amongst researchers. Thresholding can be of two types – Bi-level and Multi-level. In Bi-level thresholding, two values are assigned – one below the threshold level and the other above it. Sezgin and Sankur [3] categorized various thresholding techniques, based on histogram shape, clustering, entropy and object attributes. Otsu’s method [4] maximizes the values of class variances to get optimal threshold. Sahoo et al. [5] tested Otsu’s method on real images and concluded that the structural similarity and smoothness of reconstructed image is better than other methods. Processing time of the algorithm in Otsu's method was reduced after modification by Liao et al. [6]. In Abutaleb's method [7], threshold was calculated by using 2D entropy. Niblack's [8] method makes use of mean and standard deviation to follow a local approach. Hemachander et al. [9] proposed binarization scheme which maintains image continuity.

In Multilevel thresholding, different values are assigned between different ranges of threshold levels. Reddi et al. [10] implemented Otsu's method recursively to get multilevel thresholds. Ridler and Calward algorithm [11] defines one threshold by taking mean or any other parameter of complete image. This process is recursively used for the values below the threshold value and above it separately. Chang [12] obtained same number of classes as the number of peaks in the histogram by filtered the image histogram. Huang et al. [13] used Lorentz information measure to create an adaptive window based thresholding technique for uneven lightning of gray images. Boukharouba et al. [14] used the distribution function of the image to get multi-threshold values by specifying the zeros of a curvature function. For multi-threshold selection, Kittler and Illingworth [15] proposed a minimum error thresholding method. Papamarkos and Gatos [16] used hill clustering technique to get multi-threshold values which estimate the histogram segments by taking the global minima of rational functions. Comparison of various meta-heuristic techniques such as genetic algorithm, particle swarm optimization and differential evolution for multilevel thresholding is done by Hammouche et al. [17].

Wavelet transform has become a significant tool in the field of image processing in recent years [18][19]. Wavelet transform of an image gives four components of the image – Approximation, Horizontal, Vertical and Diagonal [20]. To match the matrix dimension of the original image, the coefficients of image is down sampled by two in both horizontal and vertical directions. To decompose image further, wavelet transform of approximation component is taken. This can continue till there is only one coefficient left in approximation part [21]. In image processing, Discrete Wavelet Transform (DWT) is widely used in compression, segmentation and multi-resolution of image [22].

In this paper a hybrid multilevel color image segmentation algorithm has been proposed, using mean and standard deviation in the wavelet domain. The method takes into account that majority of wavelet coefficients lie near to zero and coefficients representing large differences are a few in number lying at the extreme ends of histogram. Hence, the procedure provides for variable size segmentation, with bigger block size around the weighted mean, and having smaller blocks at the ends of histogram plot of each horizontal, vertical and diagonal components. For the approximation coefficients, values around weighted mean of histogram carry more information while end values of histogram are less significant. Hence, in approximation components segmentation is done with finer block size around weight mean and larger block size at the end of the histogram [1]. The algorithm is based on the fact that a number of distributions tends toward a delta function in the limit of vanishing variance. A well-known example is normal distribution

$$\lim_{\sigma \to 0} f(x) = \lim_{\sigma \to 0} \frac{1}{\sigma\sqrt{2\pi}} exp\left[-\frac{(x-\mu)^2}{2\sigma^2}\right] = \delta(x-\mu)$$

In this paper, a recently established new parameter – Structural Similarity Index Measurement (SSIM) [23] is used to compare the structural similarity of image segmented by proposed algorithm and by spatial domain algorithm with the original image. It uses mean, variance and correlation coefficient of images to relate the similarity between the images.

In section 2, illustration of approach for new hybrid algorithm is provided followed by algorithm in section 3. Section 4 consists of the observations seen and results obtained in terms of SSIM, PSNR and Time Complexity by new algorithm. Finally, section 5 provides the inference of the results obtained by the new algorithm.

## 2. METHODOLOGY

Keeping in mind the fact that wavelet transform is ideally suited for study of images because of its multi-resolution analysis ability, we implement the above principle in the wavelet domain and find that the proposed algorithm is superior to the space domain algorithm of Arora et al [1].

Following has to be done to implement the proposed methodology. Segregate the colored image $I_{RGB}$ into its Red($I_R$), Green($I_G$) and Blue($I_B$). In the proposed methodology different approaches have been applied for approximation and detail coefficients of wavelet transformed image for each $I_R$, $I_G$ and $I_B$. The coefficients are divided into blocks of variable size, using weighted mean and variance of each sub-band of histogram of coefficients. For approximation coefficients, finer block size is taken around mean while broader block size at the end of histogram. Whereas, in case of detail coefficients thresholding is done by having broader block size around mean while finer block size at the end of respective histogram. Take inverse wavelet transform for each thresholded $I_R$, $I_G$ and $I_B$ component. Reconstruct the image by concatenating $I_R$, $I_G$ and $I_B$ components. Following section provides the algorithm used.

## 3. ALGORITHM

*For Vertical/Horizontal/Diagonal coefficients*

1. Input ***n*** (no. of thresholds)
2. Input ***f*** ( Vertical/Horizontal/Diagonal coefficients matrix)
3. ***a*** = min( **f** ); ***b*** = max ( **f** );
4. $\boldsymbol{m_e}$= weighted mean ***f*** (***a*** to ***b***)
5. ***T1*** = $\boldsymbol{m_e}$ ; ***T2*** = $\boldsymbol{m_e}$ + 0.0001 ;
6. Repeat steps from (a) to (h) (***n***-1)/2 times
    (a) ***m1*** = weighted mean ***f*** (***a*** to ***T1***)
    (b) ***m2*** = weighted mean ***f*** (***T2*** to ***b*** )
    (c) ***d1*** = standard deviation ***f*** (***a*** to ***T1*** )

(d) ***d2*** = standard deviation ***f*** (***T2*** to ***b*** )

(e) ***T11*** = ***m1*** **–** (***$k_1$*** * ***d1*** ); ***T22*** = ***m2*** +( ***$k_2$*** * ***d2*** ) ;

(f) ***f*** (***T11*** to ***T1*** )= weighted mean ***f*** (***T11*** to ***T1*** )

(g) ***f*** (***T2*** to ***T22*** )= weighted mean ***f*** (***T2*** to ***T22*** )

(h) ***T1*** = ***T11*** – 0.0001; ***T2*** = ***T22*** + 0.0001;

9. ***f*** (***a*** to ***T1***)= weighted mean ***f*** (***a*** to ***T1***)

10. ***f*** (***T1*** to ***b***)= weighted mean ***f*** (***T1*** to ***b***)

11.Output ***f*** (Quantized input matrix)

*For Approximation coefficients*

1. Input ***n*** (no. of thresholds)

2. Input ***f*** ( Approximation coefficients matrix)

3. ***a*** = min( ***f*** ); ***b*** = max ( ***f*** );

4. ***$m_e$***= weighted mean ***f*** ( ***a*** to ***b*** )

5. Repeat steps from (a) to (f) (***n***-1)/2 times

(a) ***m*** = weighted mean ***f*** (***a*** to ***b***)

(b) ***d*** = standard deviation ***f*** (***a*** to ***b***)

(c) ***T1*** = ***m*** **–** (***$k_1$*** * ***d*** ); ***T2*** = ***m*** +( ***$k_2$*** * ***d*** ) ;

(d) ***f*** (***a*** to ***T1***)= weighted mean ***f*** (***a*** to ***T1***)

(e) ***f*** (***T2*** to ***b***)= weighted mean ***f*** (***T2*** to ***b***)

(f) ***a*** = ***T1*** + 0.0001; ***b*** =***T2*** **–** 0.0001;

6. $\boldsymbol{f}$ ($\boldsymbol{a}$ to $\boldsymbol{m_e}$)= weighted mean $\boldsymbol{f}$ ($\boldsymbol{a}$ to $\boldsymbol{m_e}$)

7. $\boldsymbol{f}$ ($\boldsymbol{m_e}$+1 to $\boldsymbol{b}$)= weighted mean $\boldsymbol{f}$ ($\boldsymbol{m_e}$+1 to $\boldsymbol{b}$)

8. Output $\boldsymbol{f}$ (Quantized input matrix)

## 4. RESULTS AND OBSERVATIONS

Experiments have been performed on various images using MATLAB 7.1 on a system having processing speed of 1.73 GHz and 2GB RAM. The histogram plot shown in Fig. 1 verifies the variable segmentation with bigger block size around the mean while smaller blocks at the each end of histogram plot.

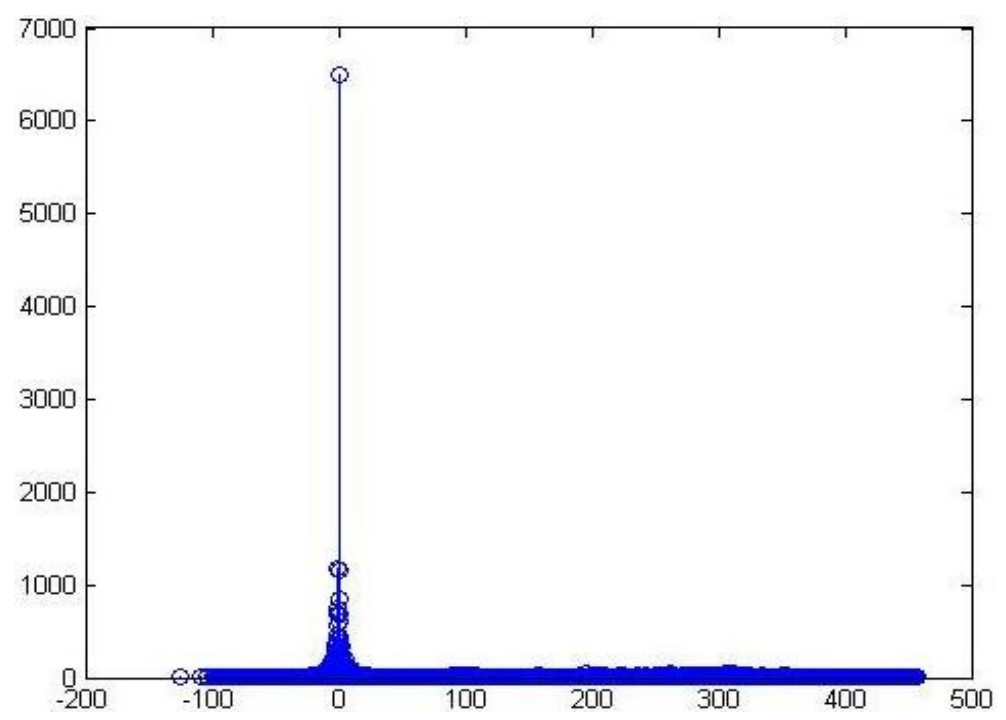

Fig.1. (a)

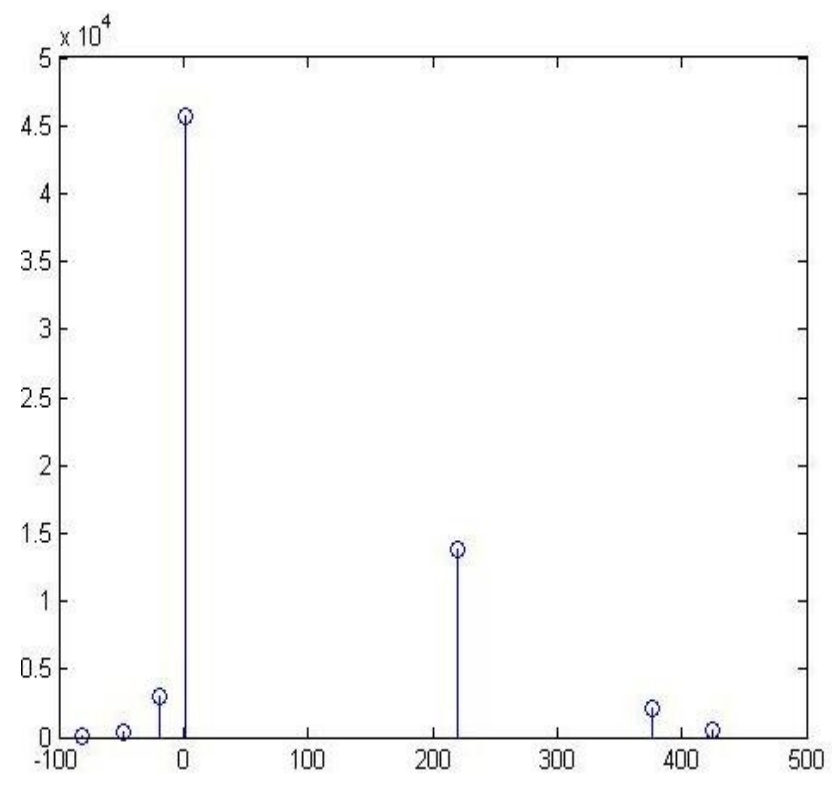

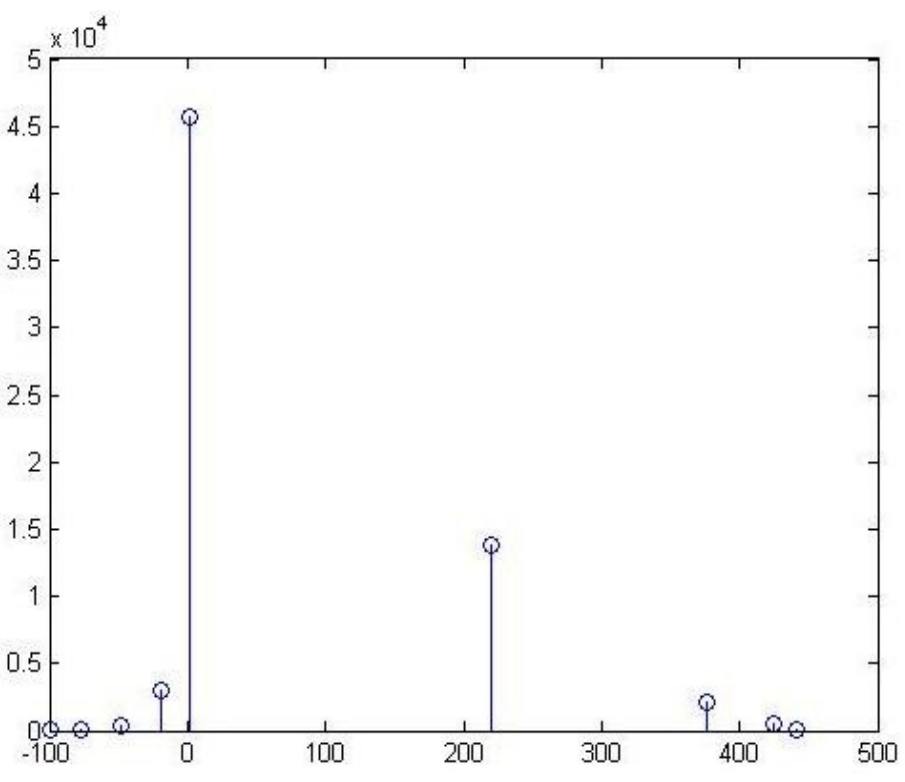

Fig.1.(b) Fig.1.(c)

Fig.1. Results: (a) histogram in wavelet domain

(b) segmentation with threshold levels seven

(c) segmentation with threshold levels nine

From the plots in Fig. 1(b) and (c), it can be easily seen that when threshold levels are increased, quantization becomes finer around the ends of histogram plot. To vary the block size, one can choose the values of $\boldsymbol{k_1}$ and $\boldsymbol{k_2}$ accordingly. The result of proposed algorithm is tested on variety of images.

In Fig. 2, original Aerial image with segmented images in space domain and by proposed algorithm are shown at different thresholding levels.

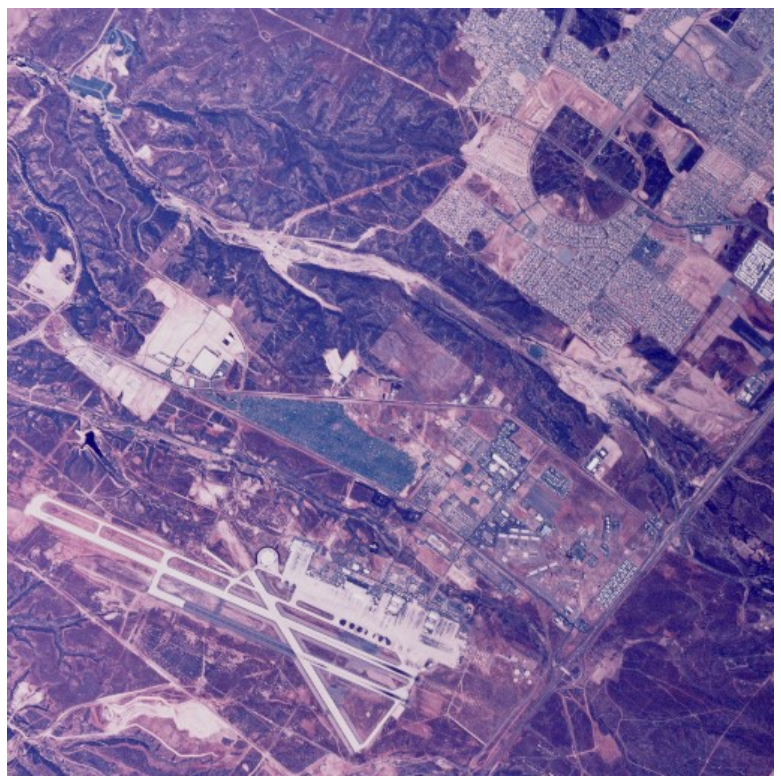

Fig.2. (a)

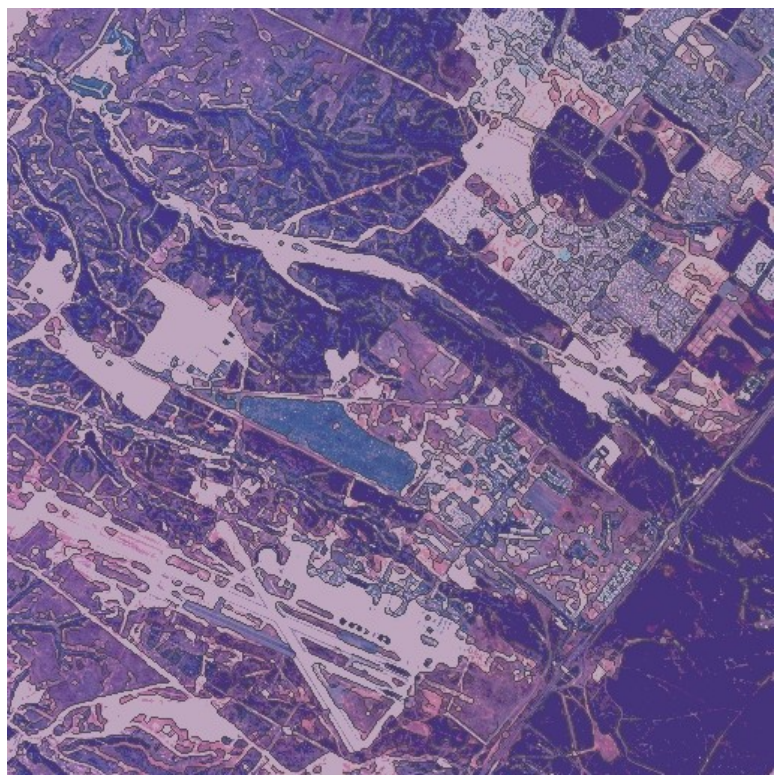

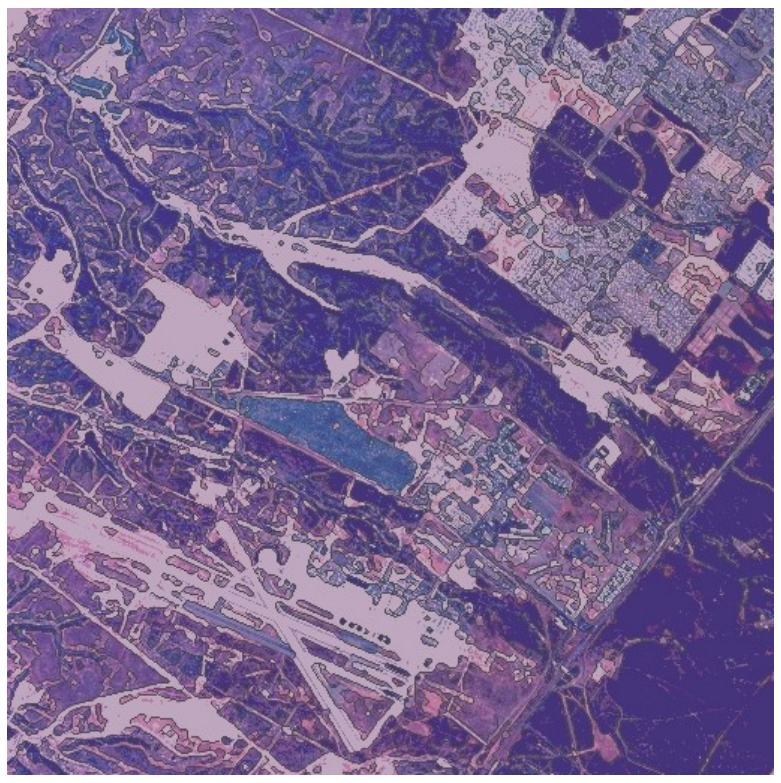

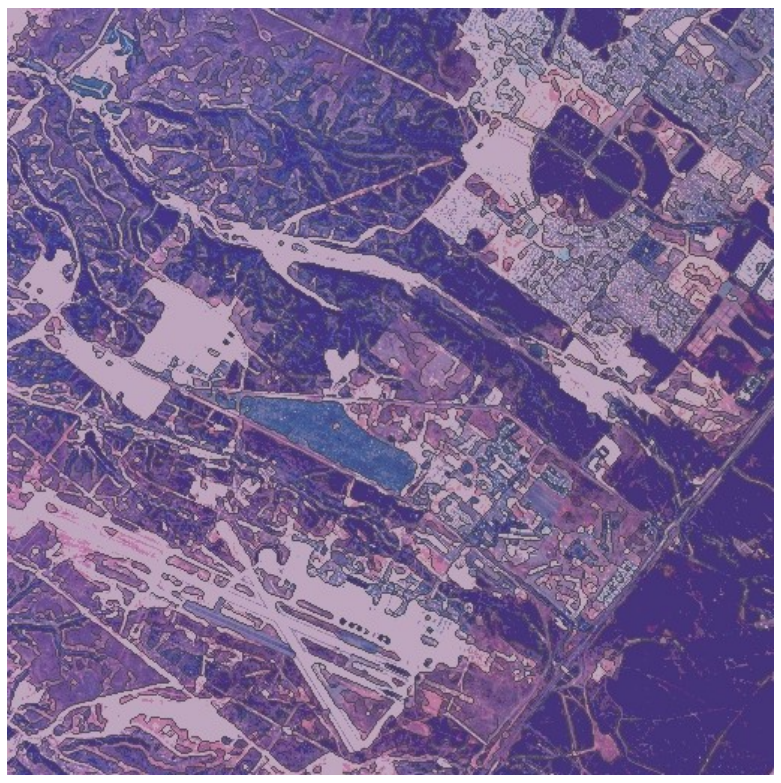

Fig.2. (b) Fig.2. (c) Fig.2. (d)

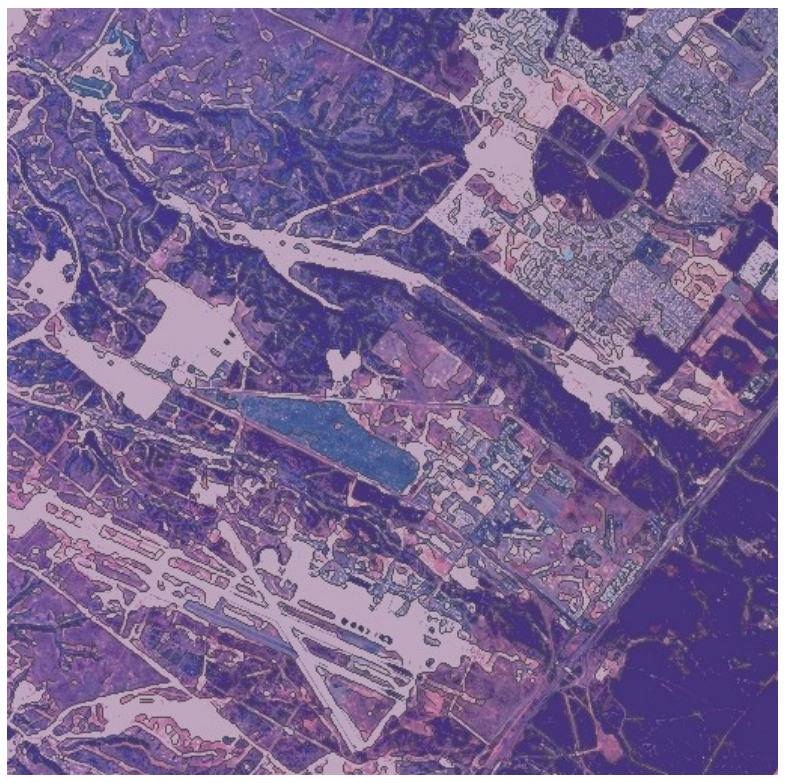

Fig.2. (e)

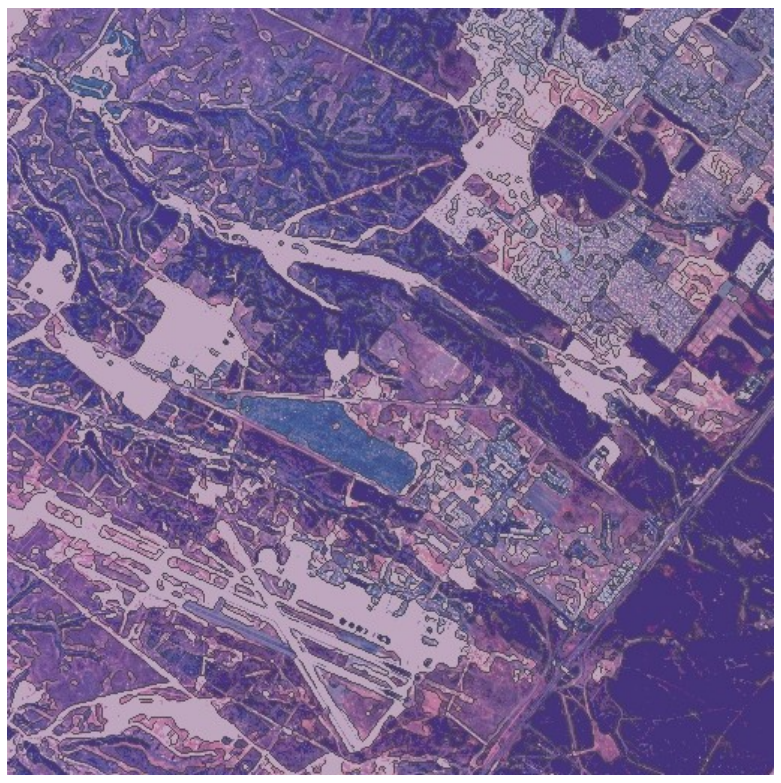

Fig.2. (f)

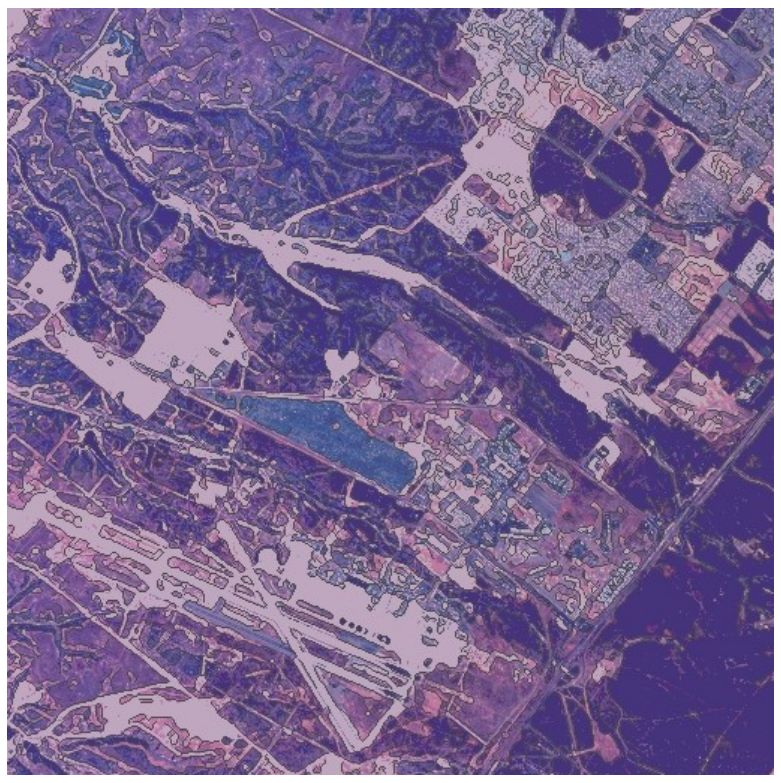

Fig.2. (g)

Fig.2. Results:(a) Original Aerial Image.

(b,c,d) Segmentation in space domain at threshold level 3,5 and 7.

(e,f,g) Segmentation by proposed algorithm at threshold level 3,5 and 7.

In Fig. 3, histograms of Approximation, Horizontal, Vertical and Diagonal coefficients of R, G and B components of original and segmented Aerial image in wavelet domain is depicted.

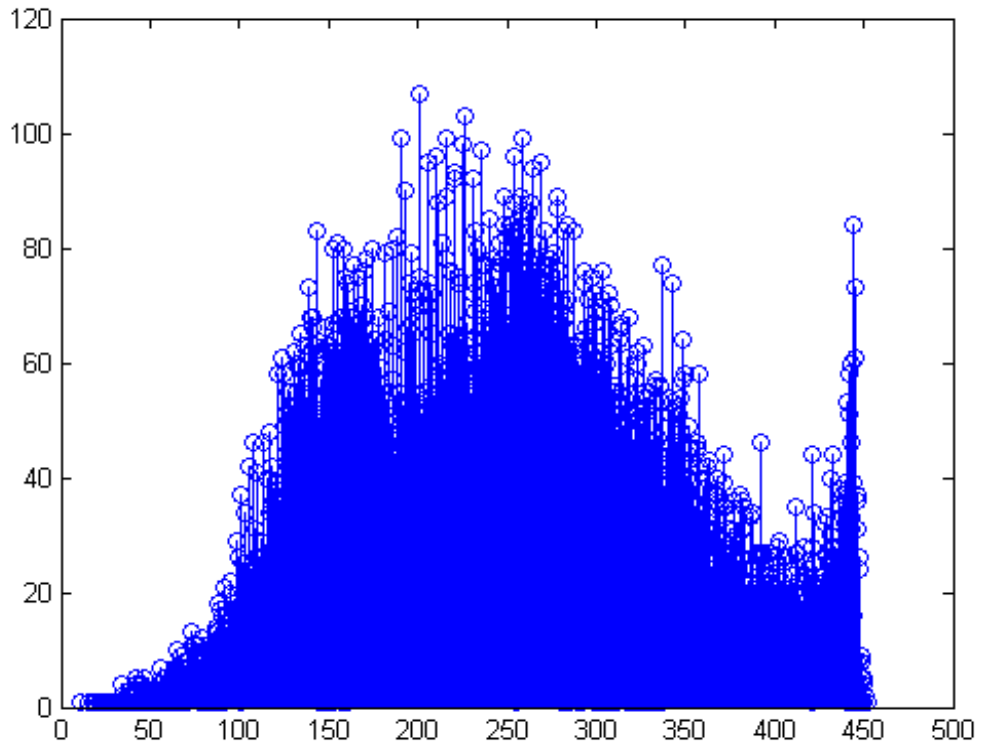

Fig.3. (1)

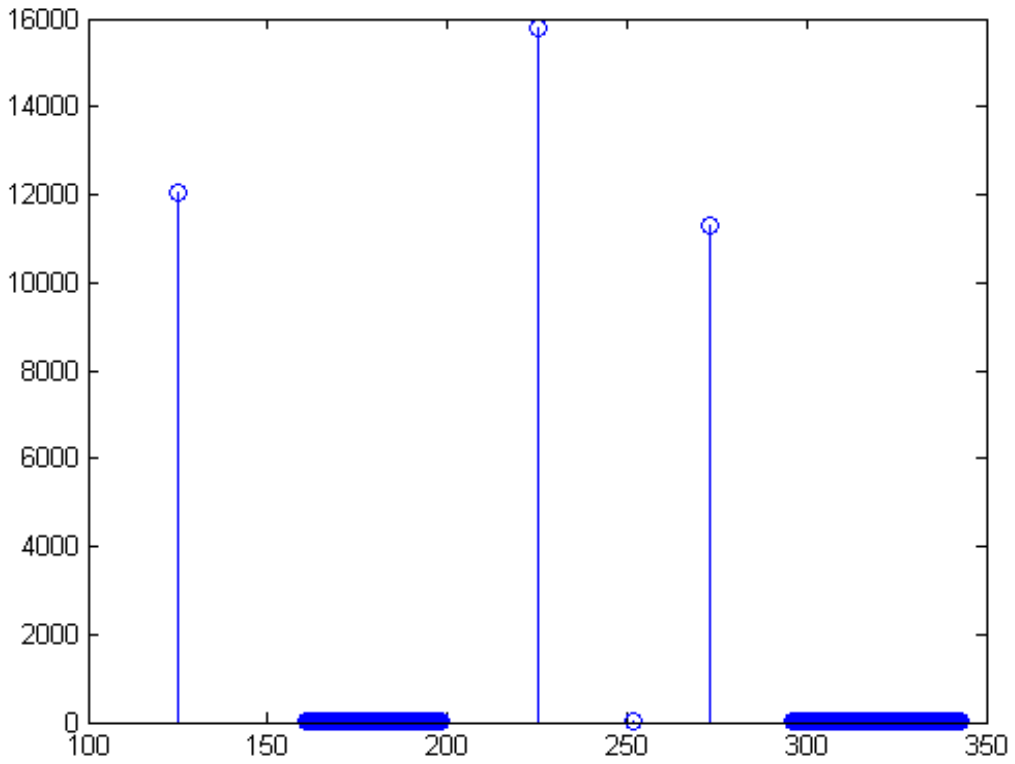

Fig.3. (2)

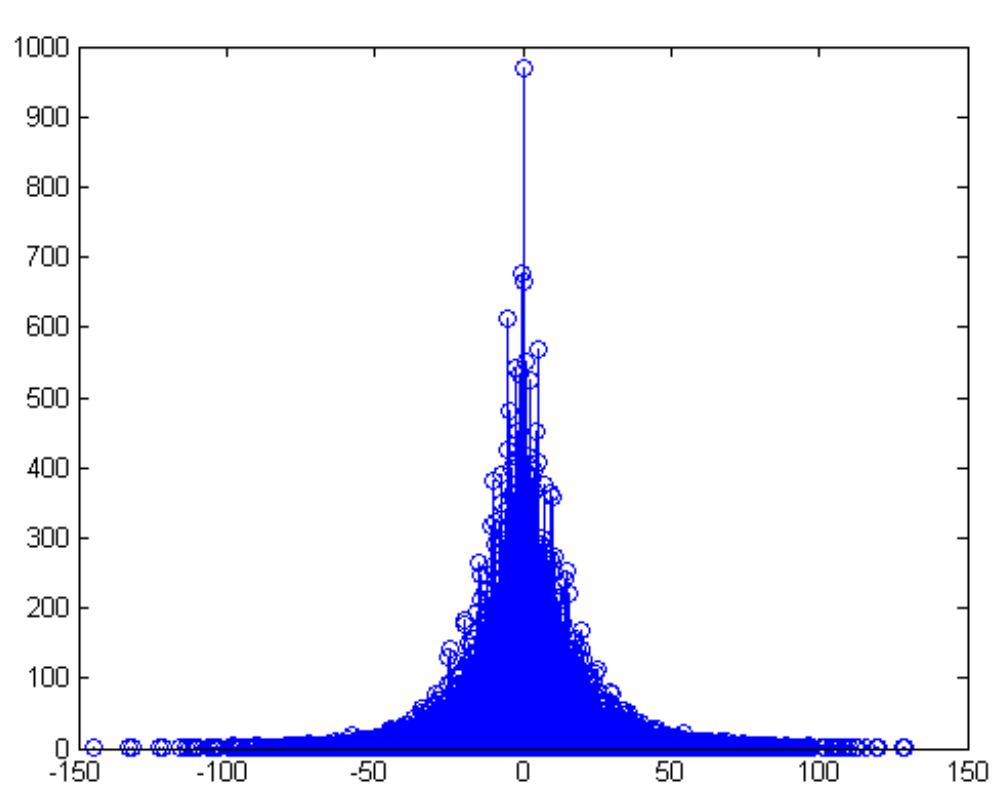

Fig.3. (3)

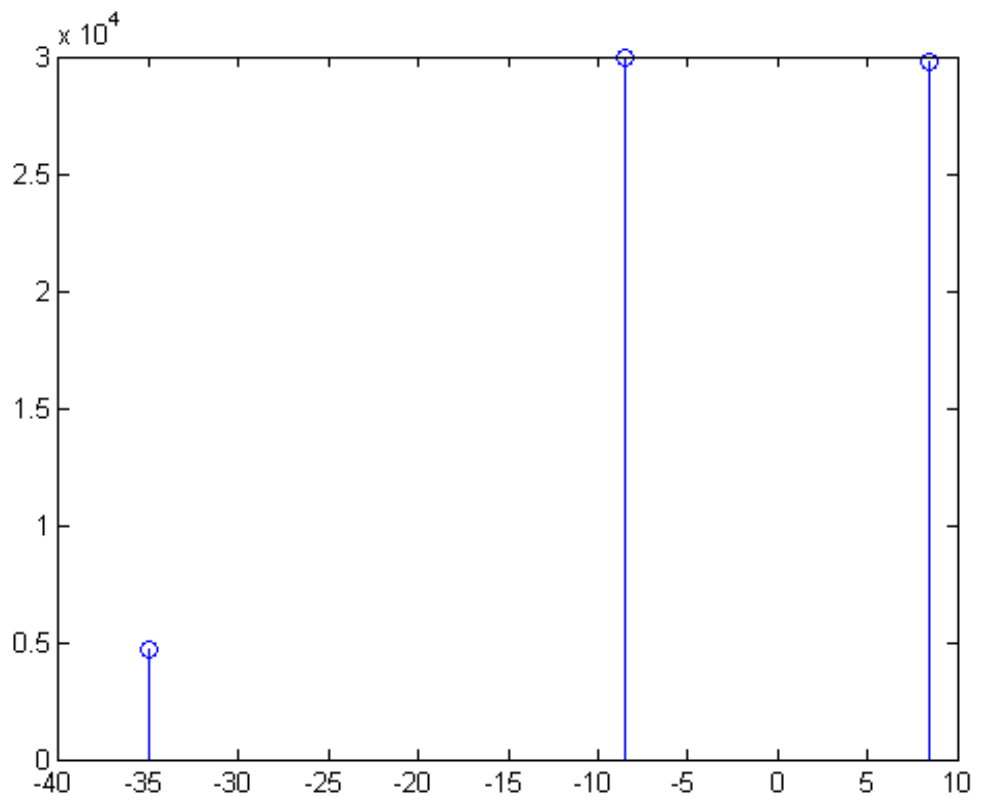

Fig.3. (4)

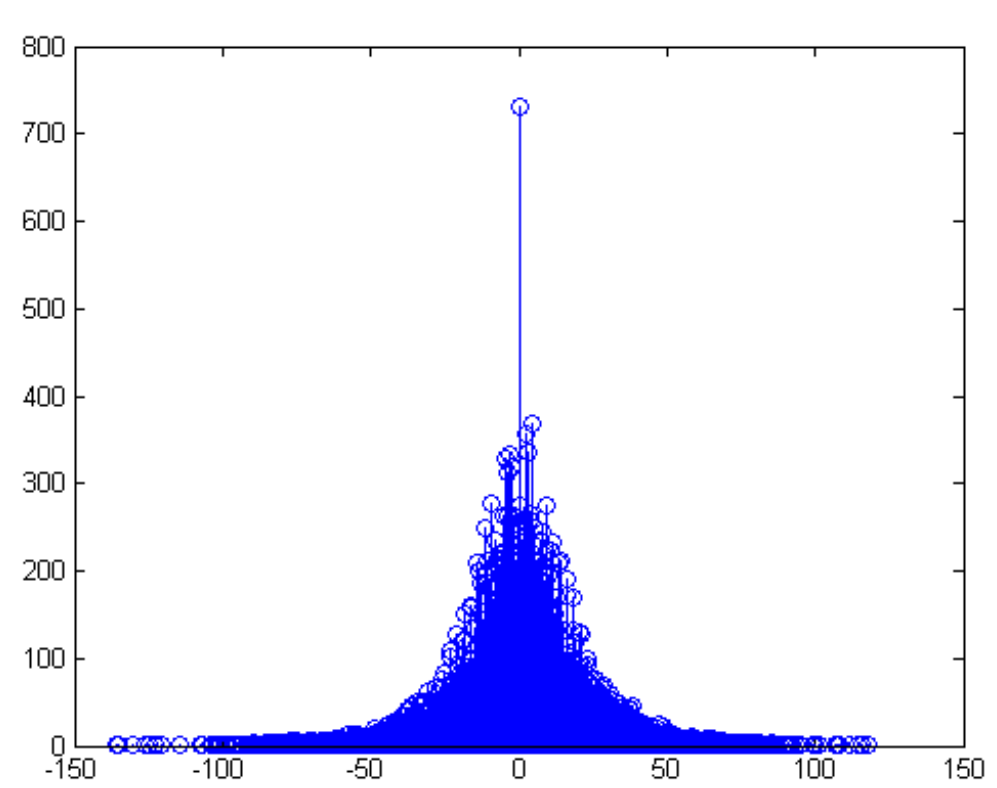

Fig.3. (5)

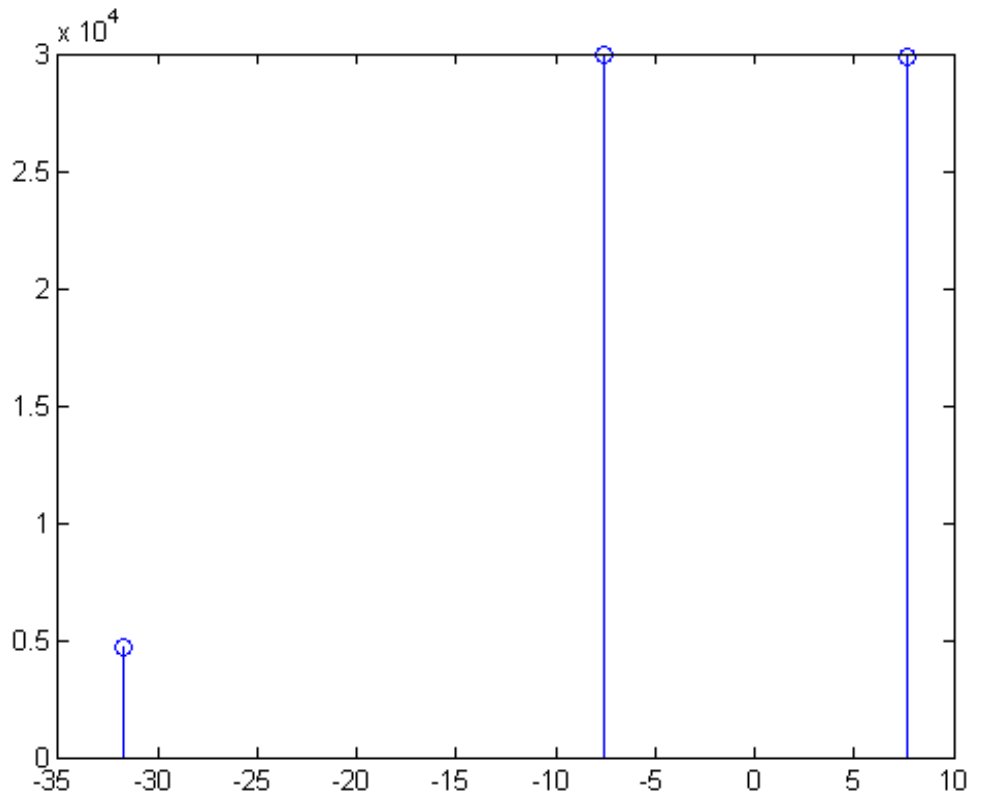

Fig.3. (6)

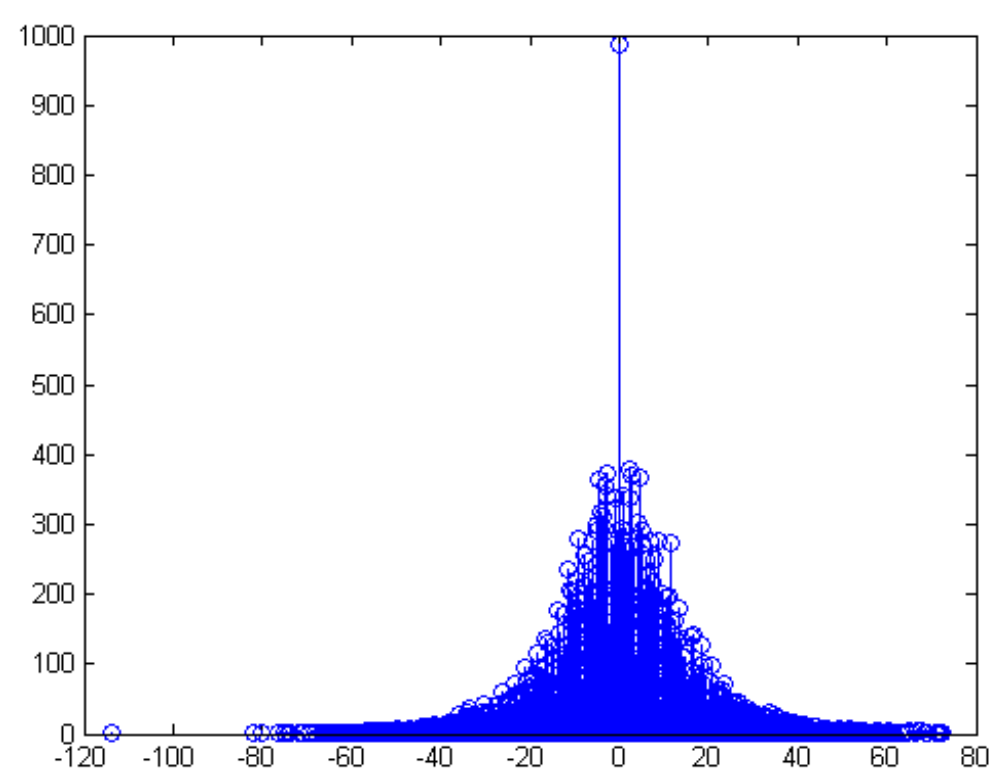

Fig.3. (7)

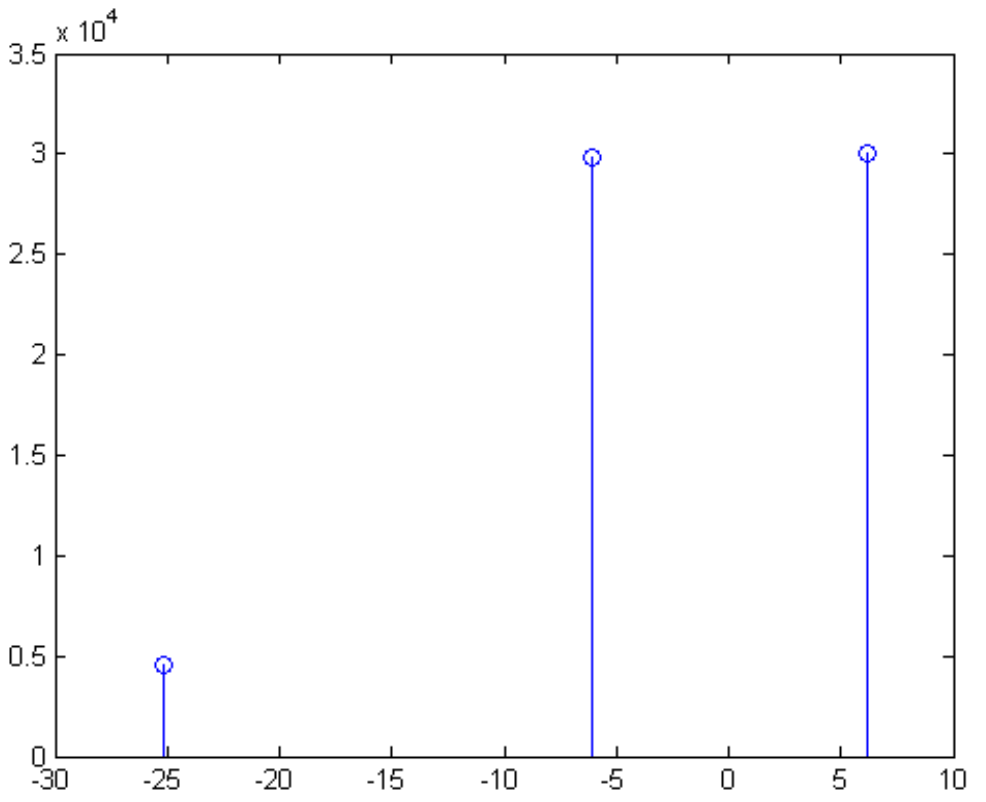

Fig.3. (8)

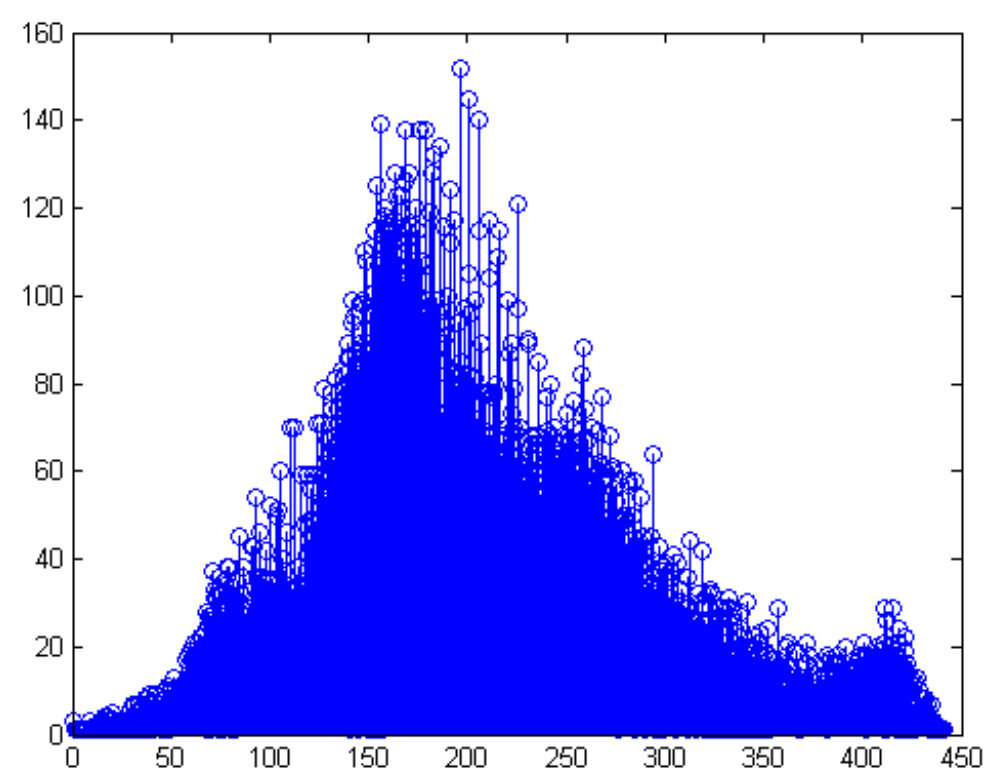

Fig.3. (9)

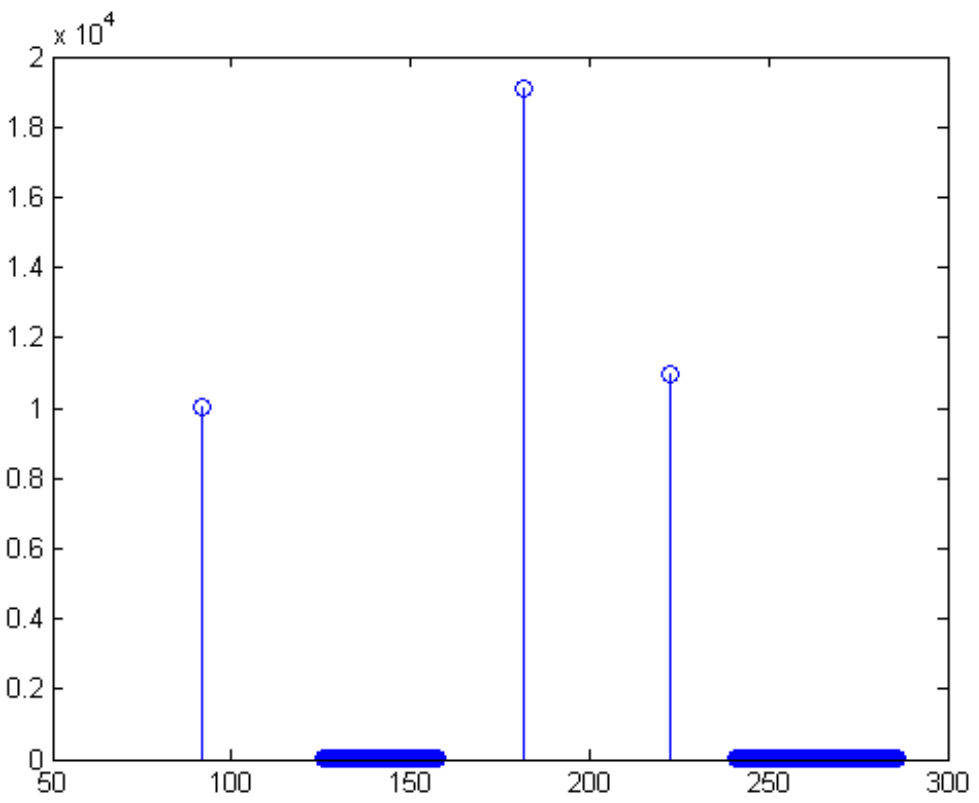

Fig.3. (10)

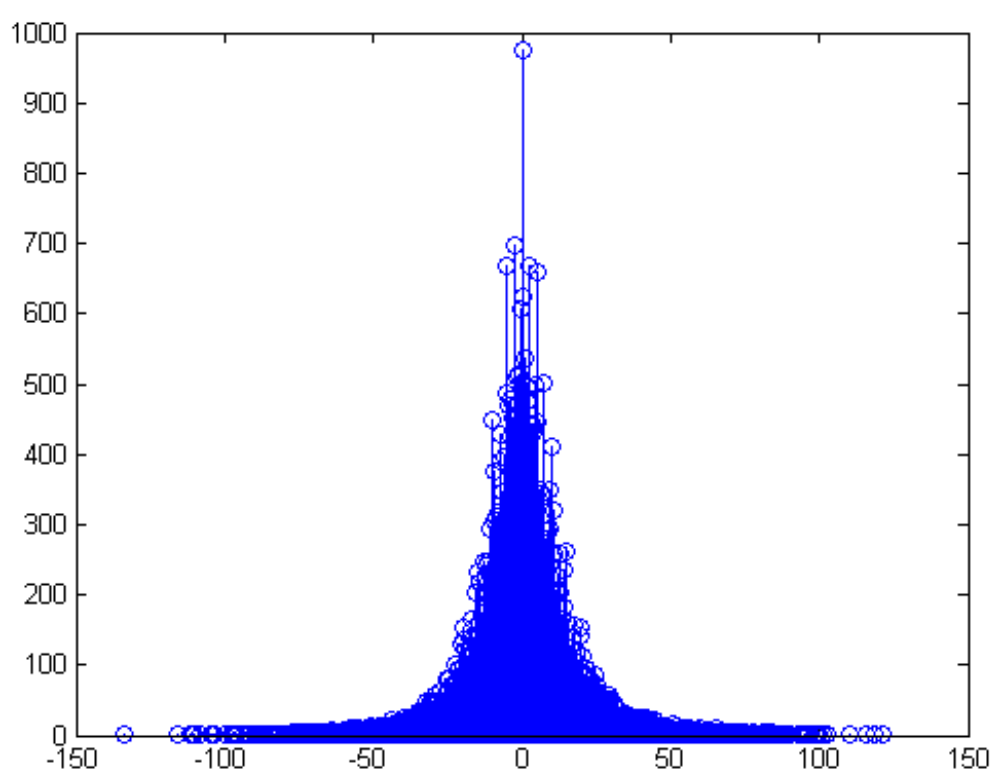

Fig.3. (11)

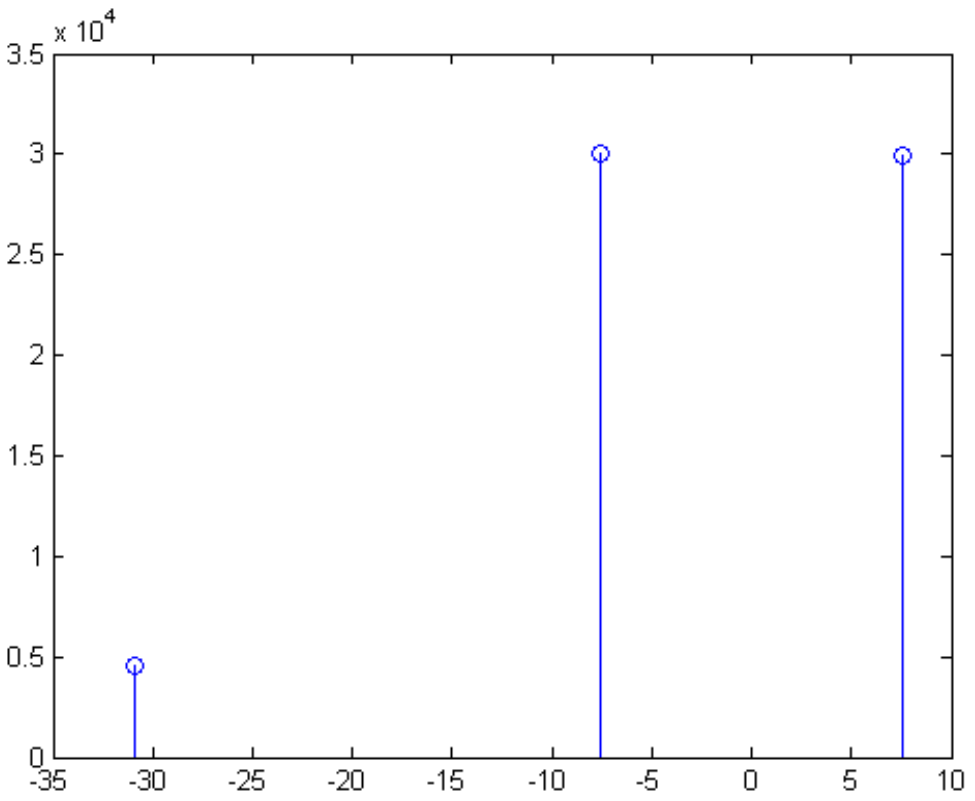

Fig.3. (12)

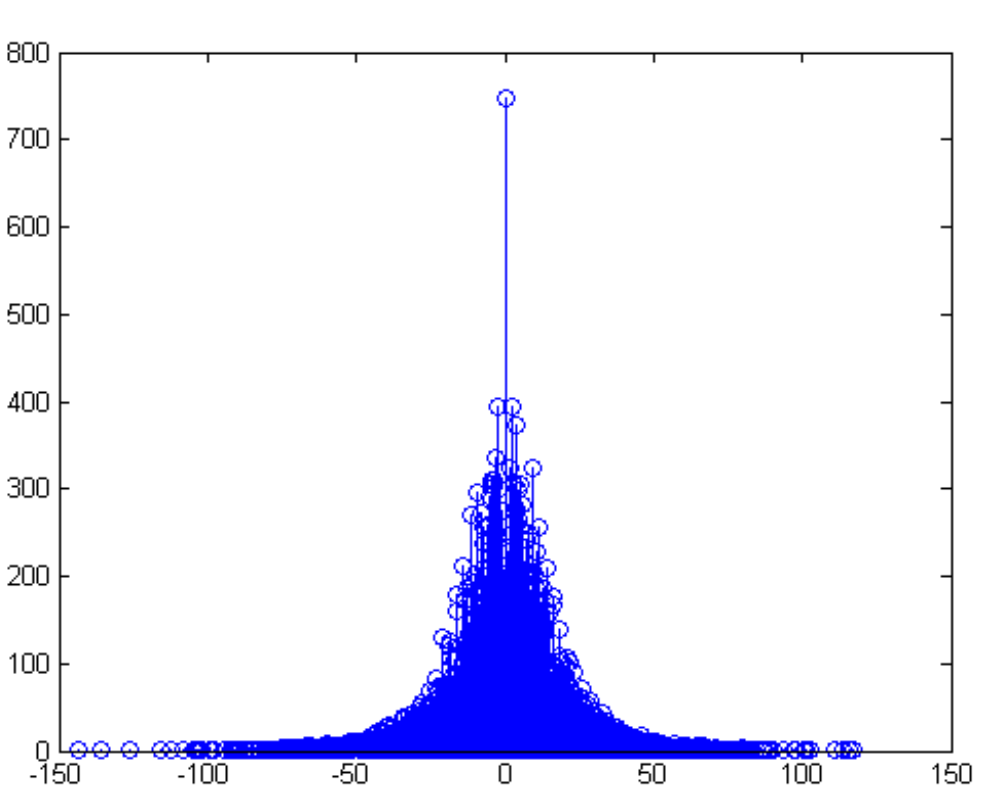

Fig.3. (13)

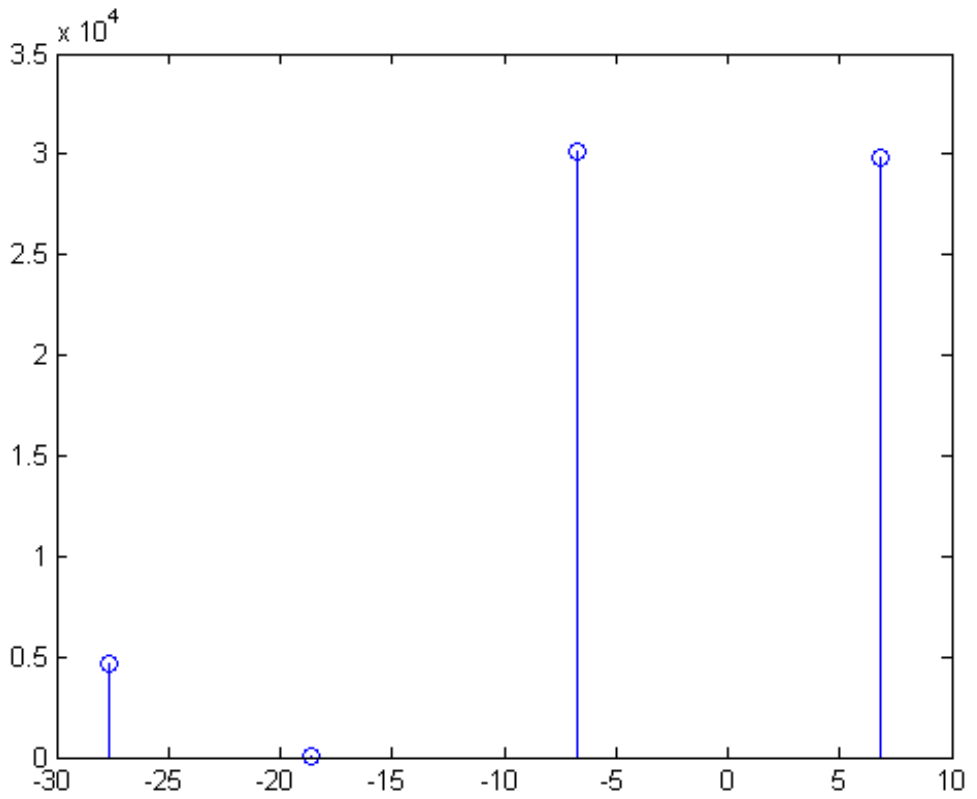

Fig.3. (14)

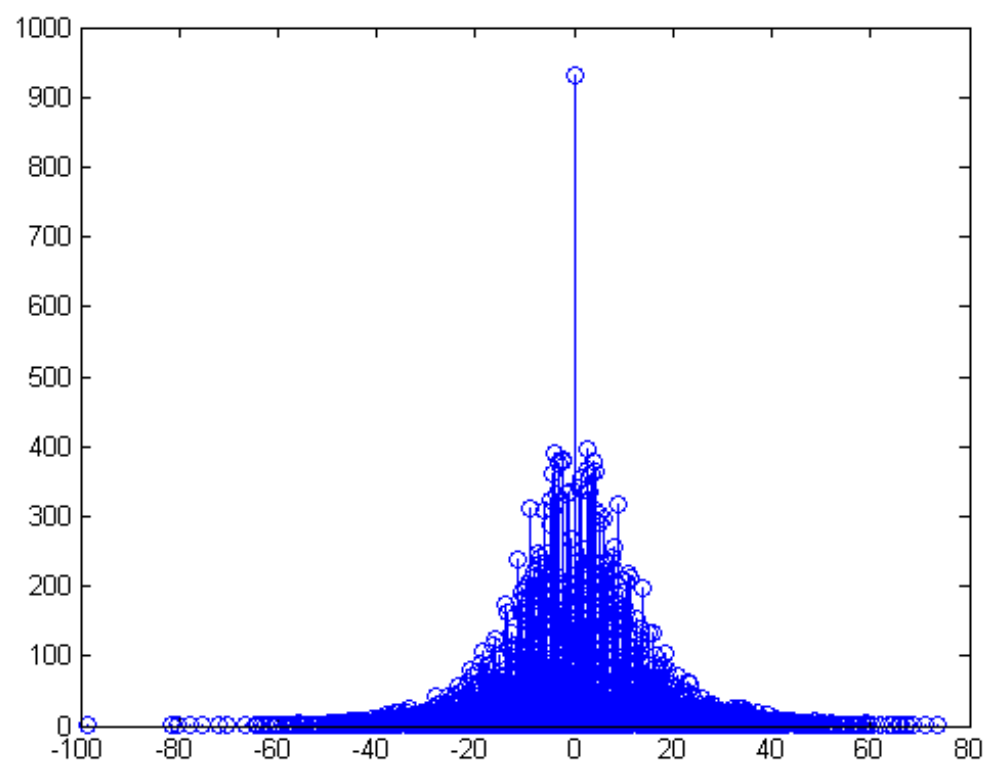

Fig.3. (15)

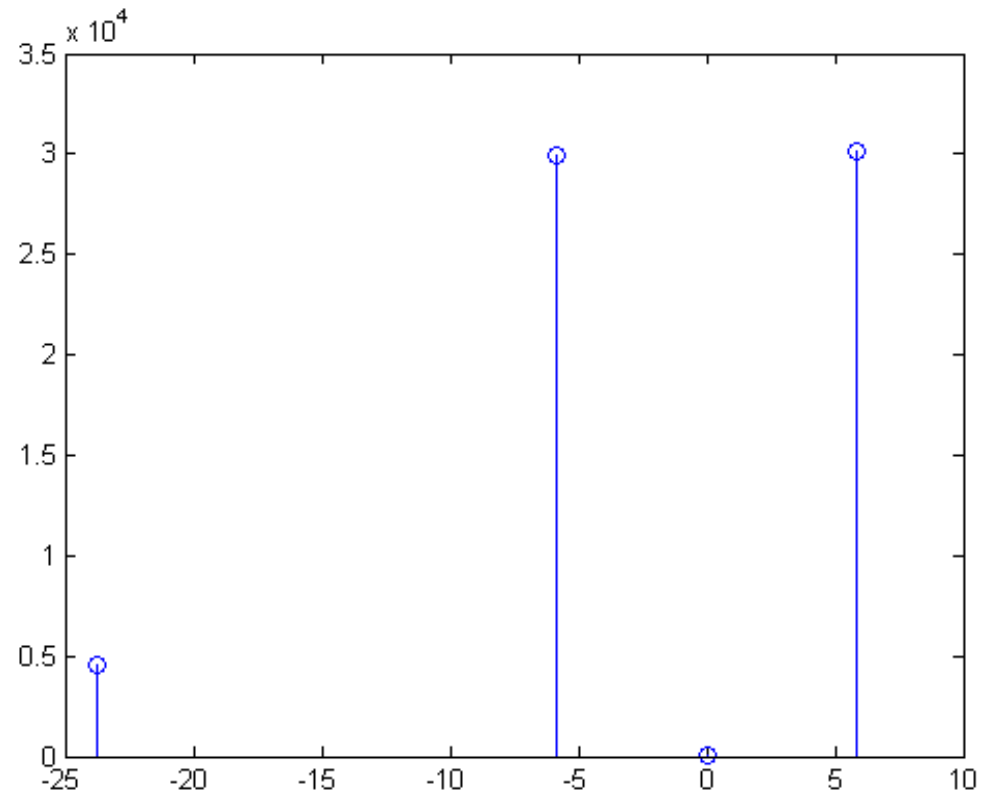

Fig.3. (16)

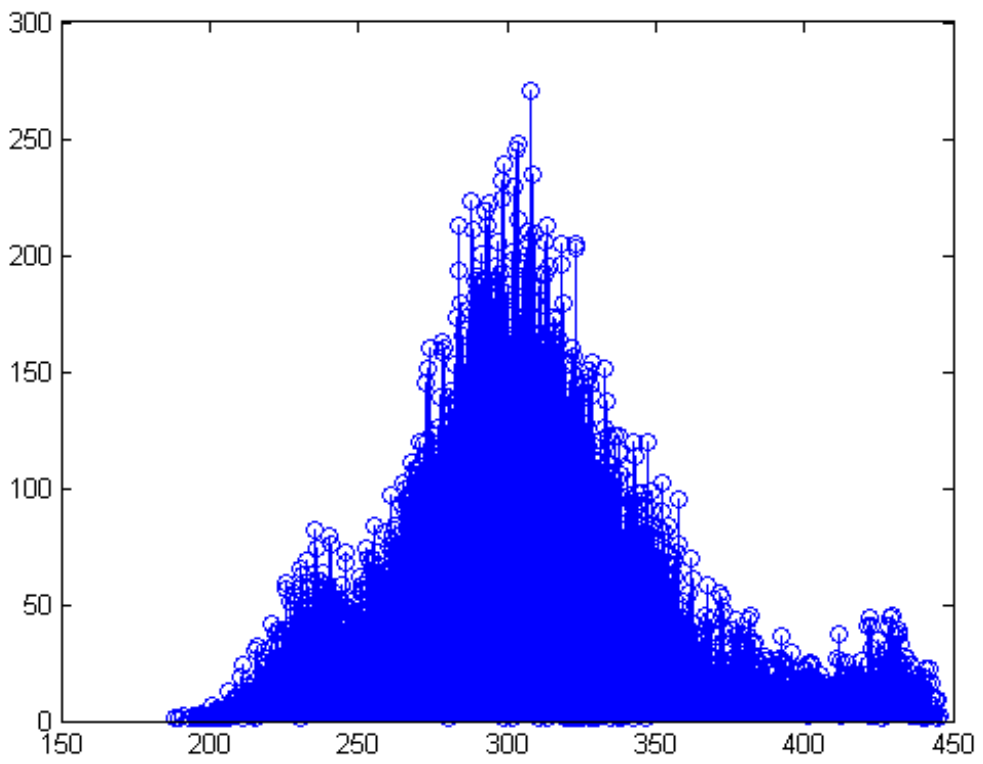

Fig.3. (17)

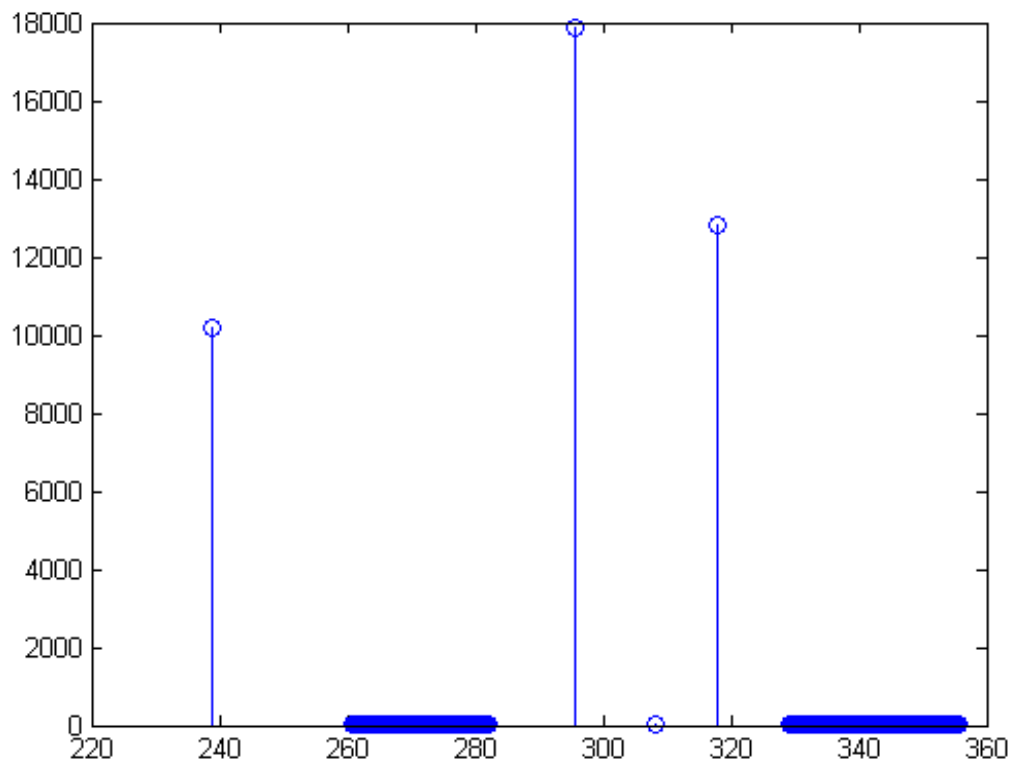

Fig.3. (18)

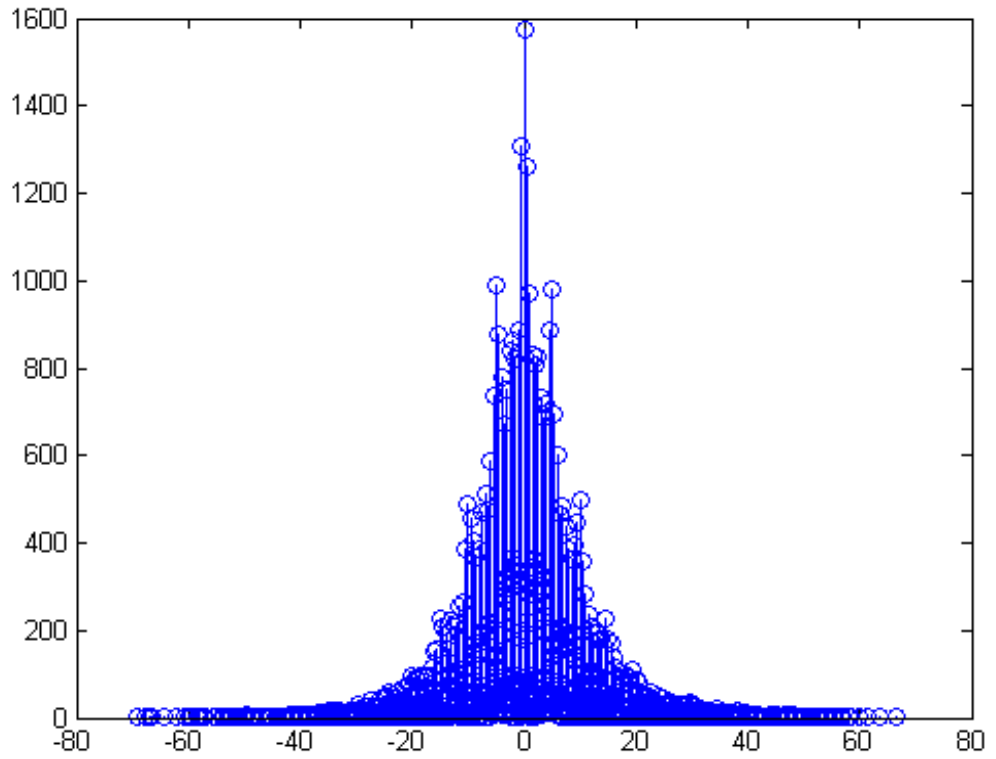

Fig.3. (19)

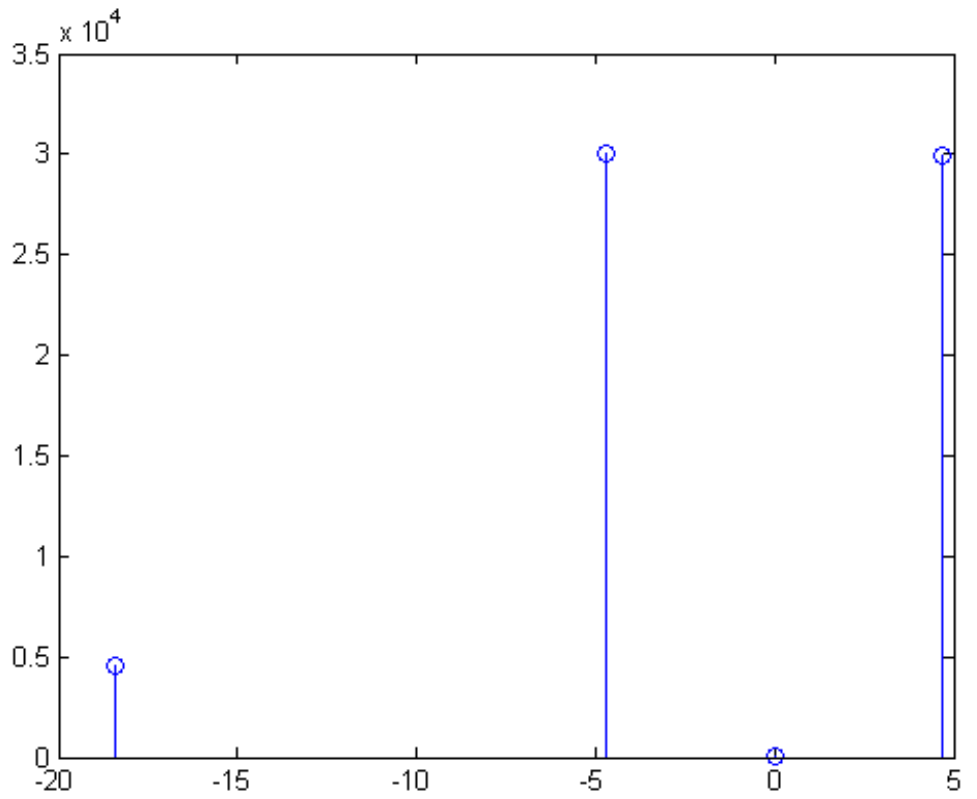

Fig.3. (20)

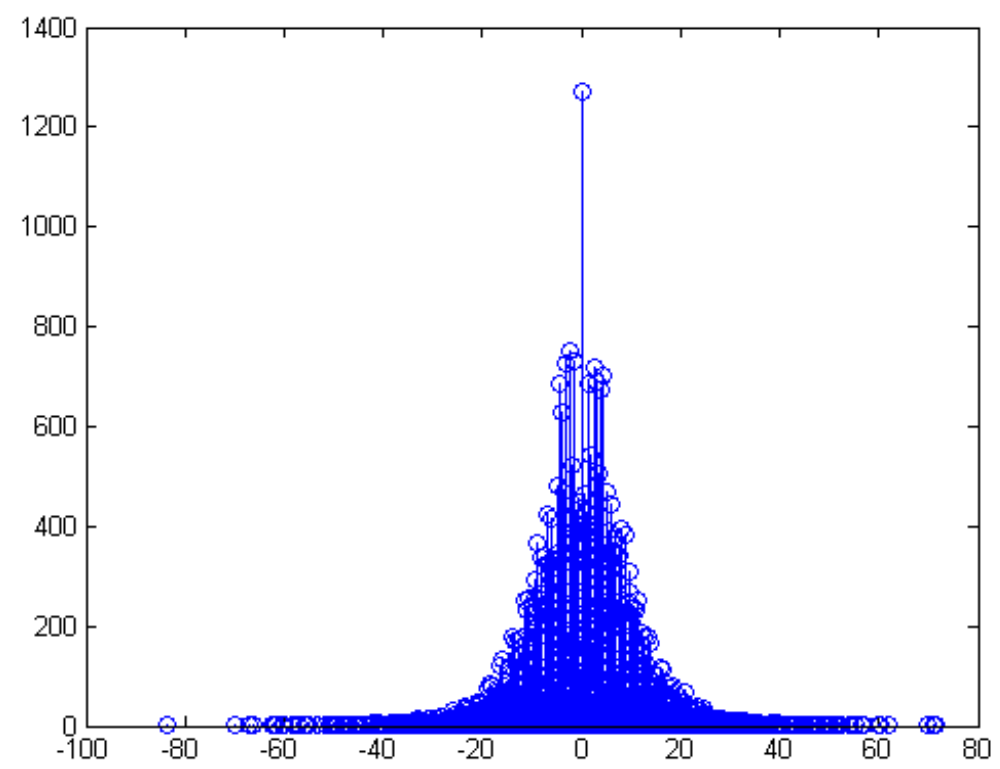

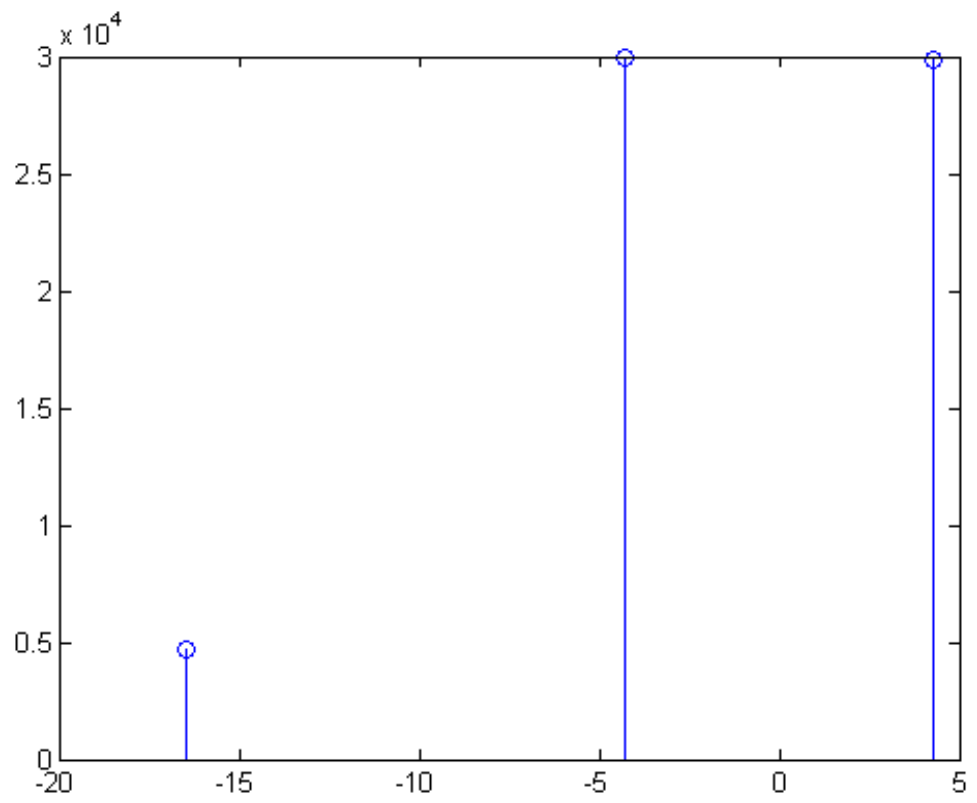

Fig.3. (21) Fig.3. (22)

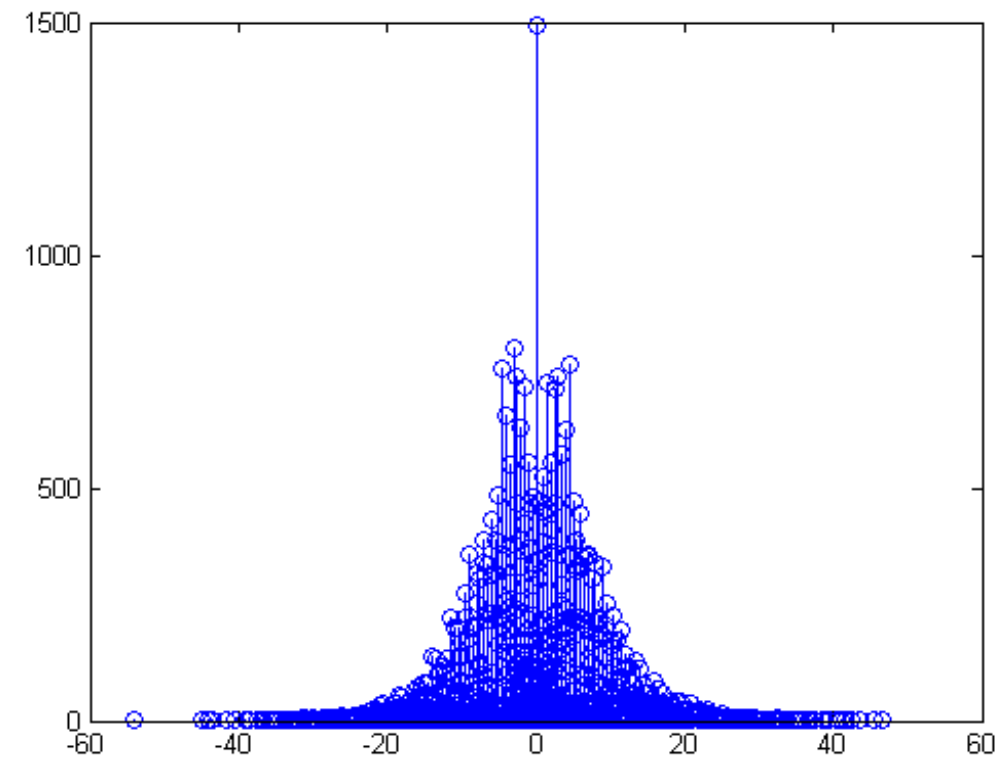

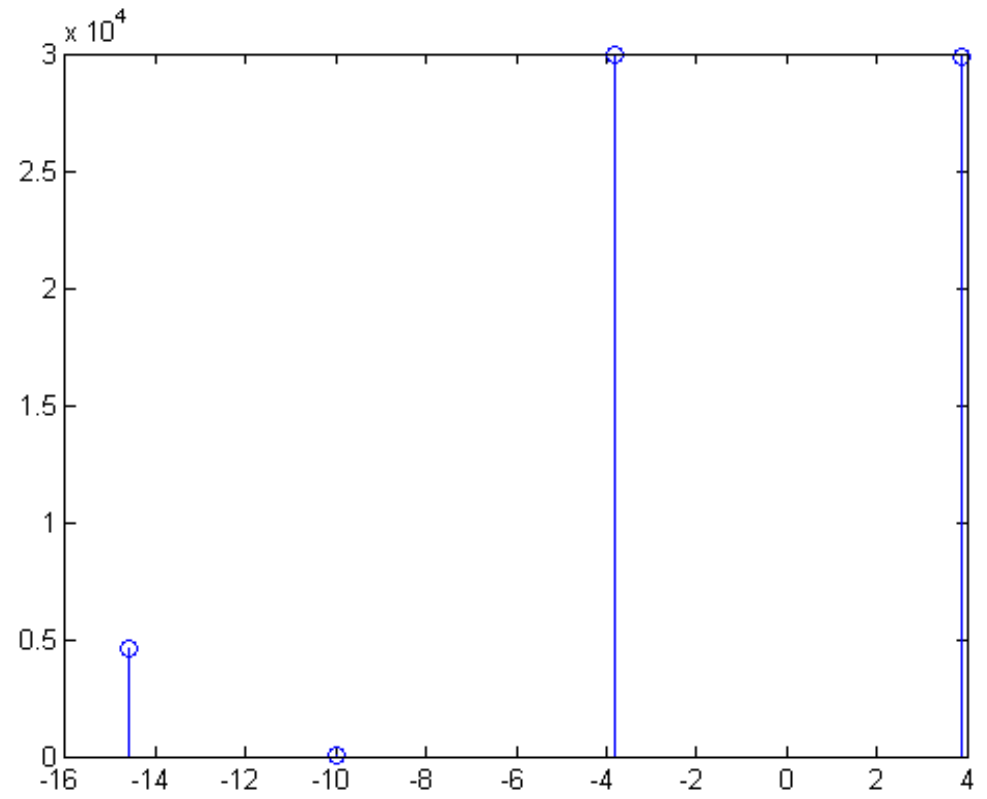

Fig.3. (23) Fig.3. (24)

Fig.3. Results:(1,9,17) Histogram of Approximation coefficients of R, G and B components respectively in wavelet domain of Aerial image.

(2,10,18) Histogram of Approximation coefficients of R, G and B components respectively in wavelet domain of Aerial image thresholded at level 3.

(3,11,19) Histogram of Horizontal coefficients of R, G and B components respectively in wavelet domain of Aerial image.

(4,12,20) Histogram of Horizontal coefficients of R, G and B components respectively in wavelet domain of Aerial image thresholded at level 3.

(5,13,21) Histogram of Vertical coefficients of R, G and B components respectively in wavelet domain of Aerial image.

(6,14,22) Histogram of Vertical coefficients of R, G and B components respectively in wavelet domain of Aerial image thresholded at level 3.

(7,15,23) Histogram of Diagonal coefficients of R, G and B components respectively in wavelet domain of Aerial image.

(8,16,24) Histogram of Diagonal coefficients of R, G and B components respectively in wavelet domain of Aerial image thresholded at level 3.

Fig. 4 shows the histogram of segmented R, G and B components of Aerial image in wavelet domain at thresholding levels 3, 5 and 7.

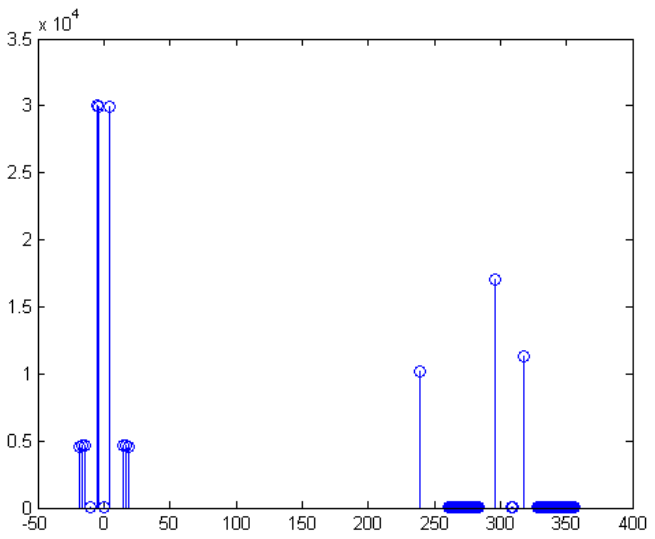

Fig.4. (a)

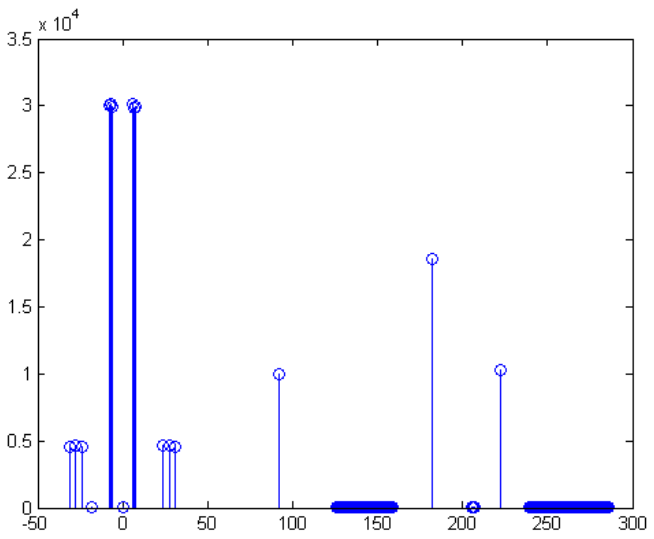

Fig.4. (b)

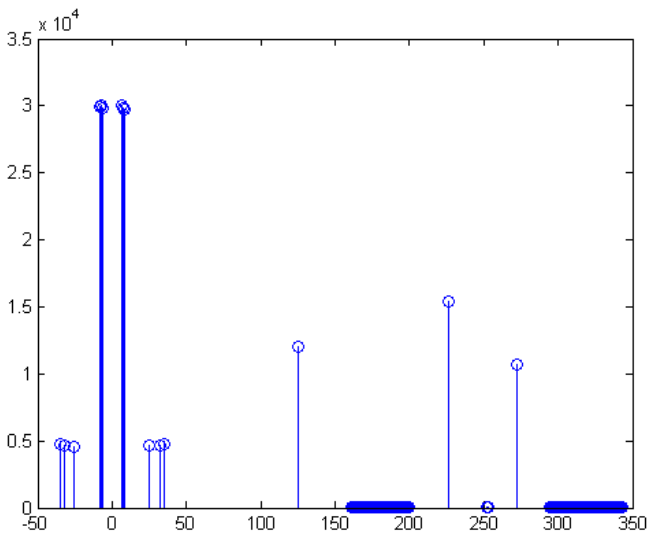

Fig.4. (c)

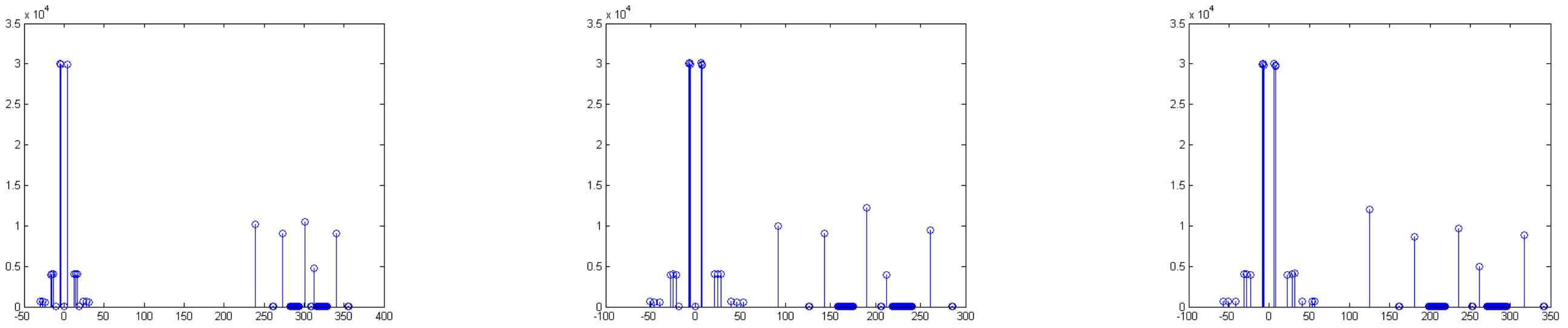

Fig.4. (d) Fig.4. (e) Fig.4. (f)

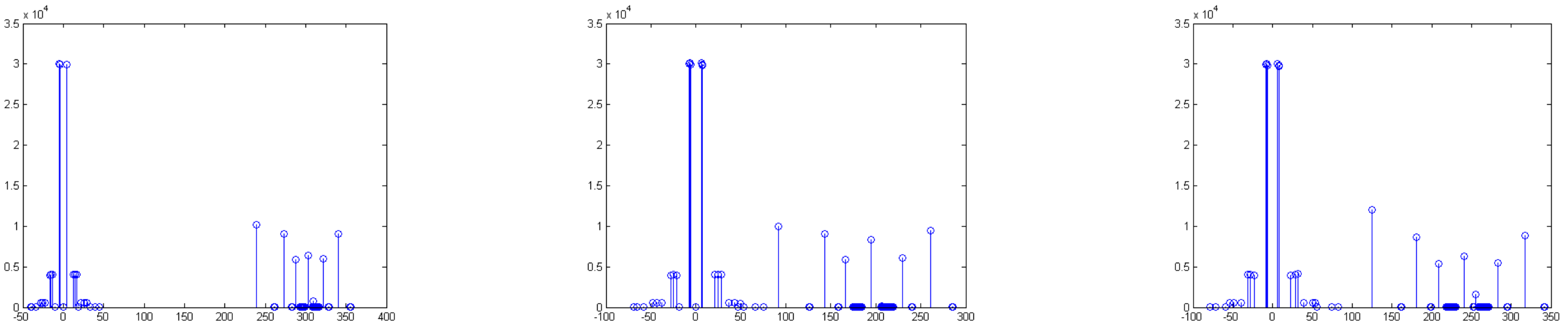

Fig.4. (g) Fig.4. (h) Fig.4. (i)

Fig.4. Results:(a,b,c) Histogram of B, G and R components thresholded in wavelet domain at level 3 of Aerial image.

(d,e,f) Histogram of B, G and R components thresholded in wavelet domain at level 5 of Aerial image.

(g,h,i) Histogram of B, G and R components thresholded in wavelet domain at level 7 of Aerial image.

*Table 1: Comparison of SSIM[20] of Aerial image (512 x512, 768.1 kB) between image segmentation in space domain and by proposed algorithm.*

| Threshold Level | SSIM in Space Domain | SSIM in Wavelet Domain |
|---|---|---|
| 3 | 0.9520 | 0.9666 |
| 5 | 0.9506 | 0.9676 |
| 7 | 0.9505 | 0.9678 |

*Table 2: Comparison of PSNR of Aerial image (512 x512, 768.1 kB) between image segmentation in space domain and by proposed algorithm.*

| Threshold Level | PSNR in Space Domain (dB) | PSNR in Wavelet Domain (dB) |
|---|---|---|
| 3 | 22.1656 | 23.1395 |
| 5 | 22.0107 | 23.4560 |
| 7 | 22.0045 | 23.5226 |

*Table 3: Comparison of Time Complexity of Aerial image (512 x512, 768.1 kB) between image segmentation in space domain and by proposed algorithm.*

| Threshold Level | Time Complexity in Space Domain (sec) | Time Complexity in Wavelet Domain (sec) |
|---|---|---|
| 3 | 3.0888 | 1.6224 |
| 5 | 3.6192 | 1.7628 |
| 7 | 3.9780 | 1.9188 |

The results of Table 1, 2 and 3 are plotted in the fig. 5. The red color graph represents the result of space domain algorithm while the black color represents the results of proposed algorithm.

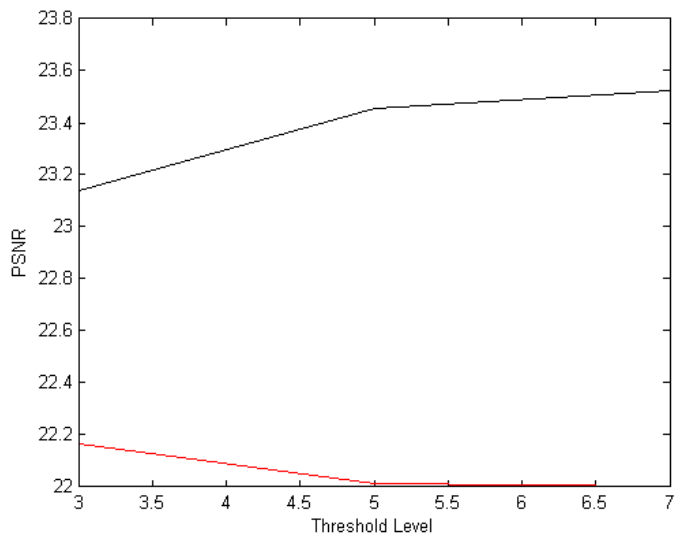

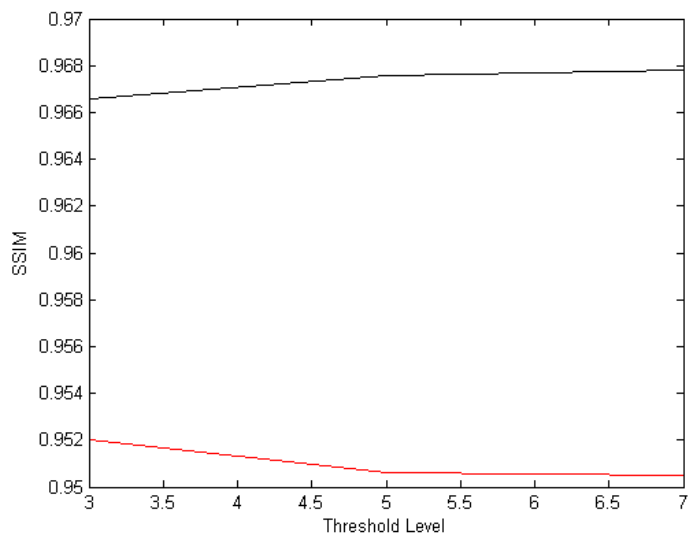

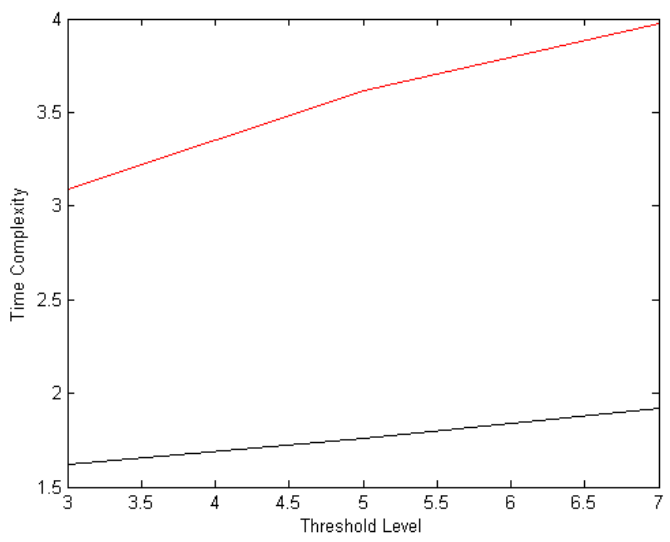

Fig.5. (a) Fig.5. (b) Fig.5. (c)

Fig.5. Results: (a) Comparison of PSNR of Aerial image between image segmentation in space domain and by proposed algorithm.

(b) Comparison of SSIM of Aerial image between image segmentation in space domain and by proposed algorithm.

(c) Comparison of Time Complexity of Aerial image between image

segmentation in space domain and by proposed algorithm.

In Fig. 6, original Earth image with segmented images in space domain and by proposed algorithm are shown at different thresholding levels.

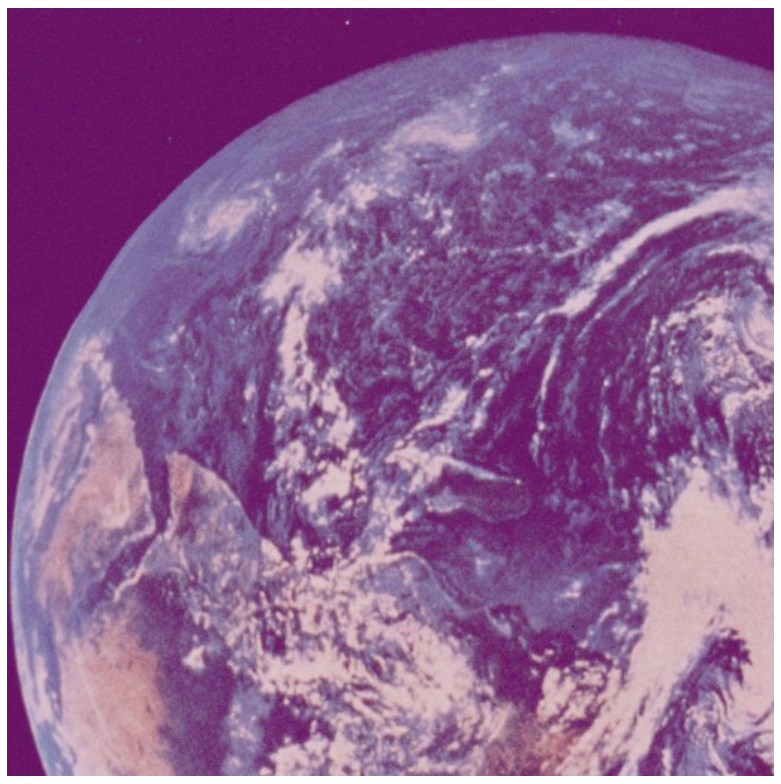

Fig.6. (a)

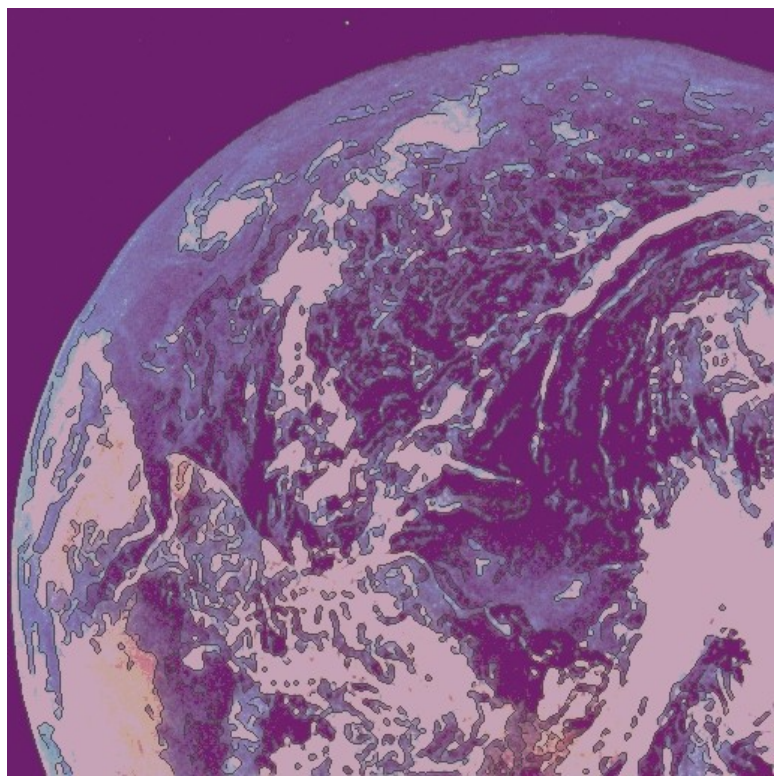

Fig.6. (b)

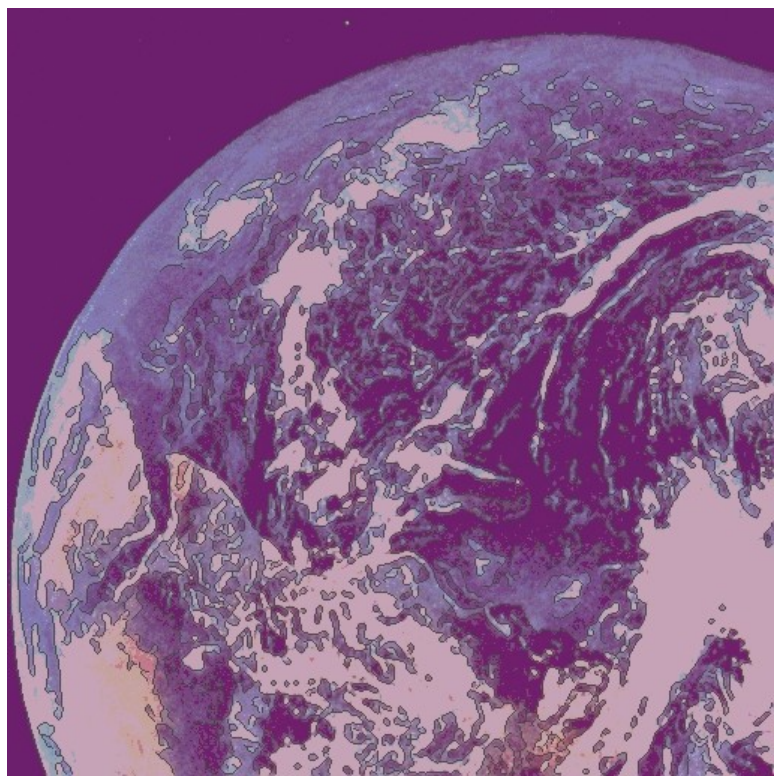

Fig.6. (c)

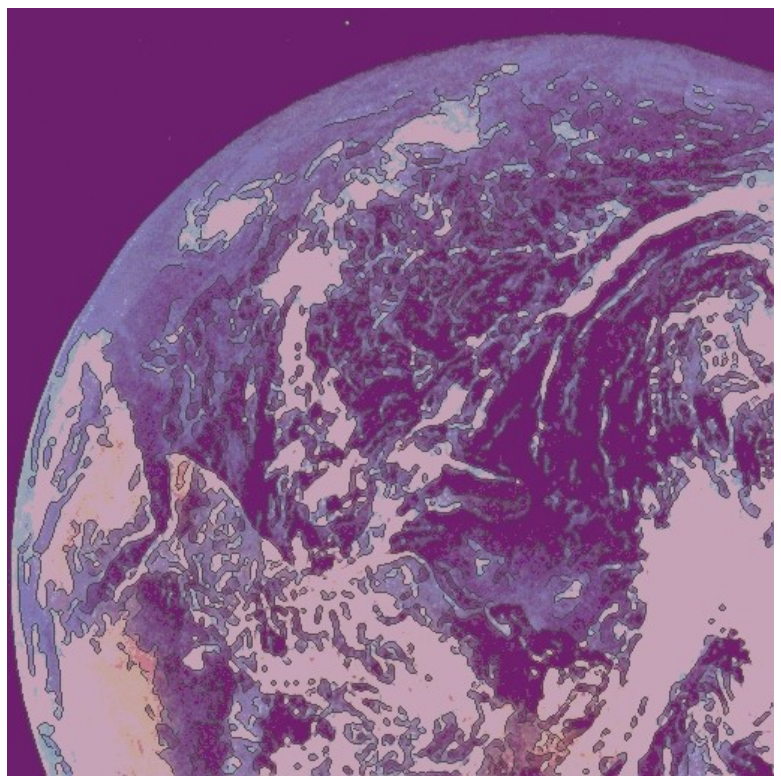

Fig.6. (d)

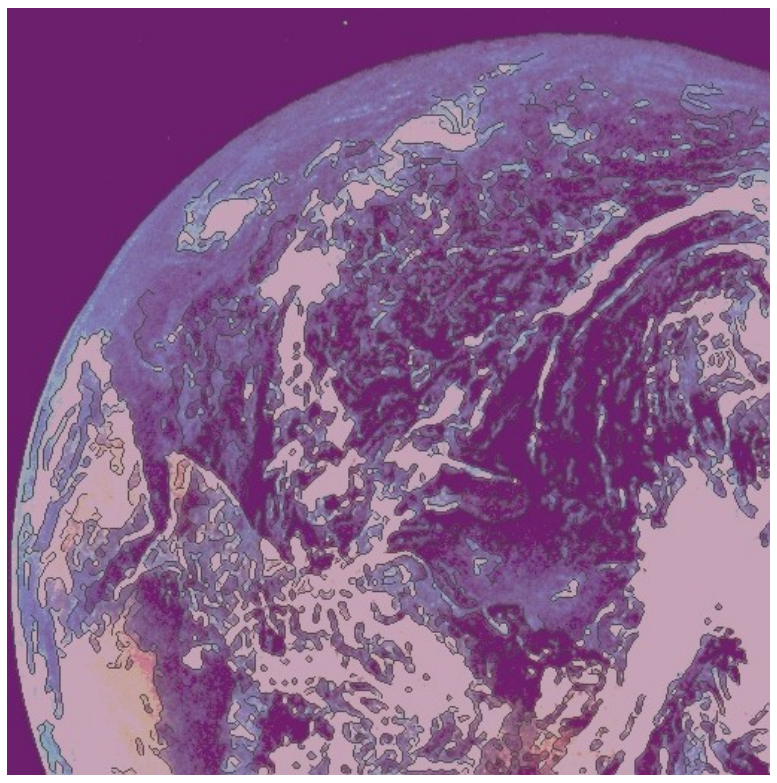

Fig.6. (e)

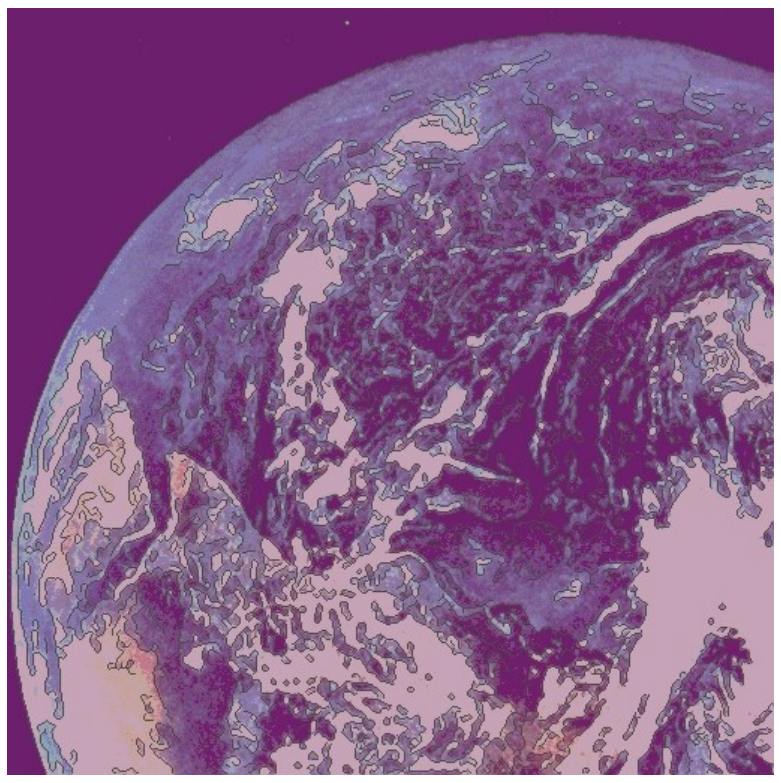

Fig.6. (f)

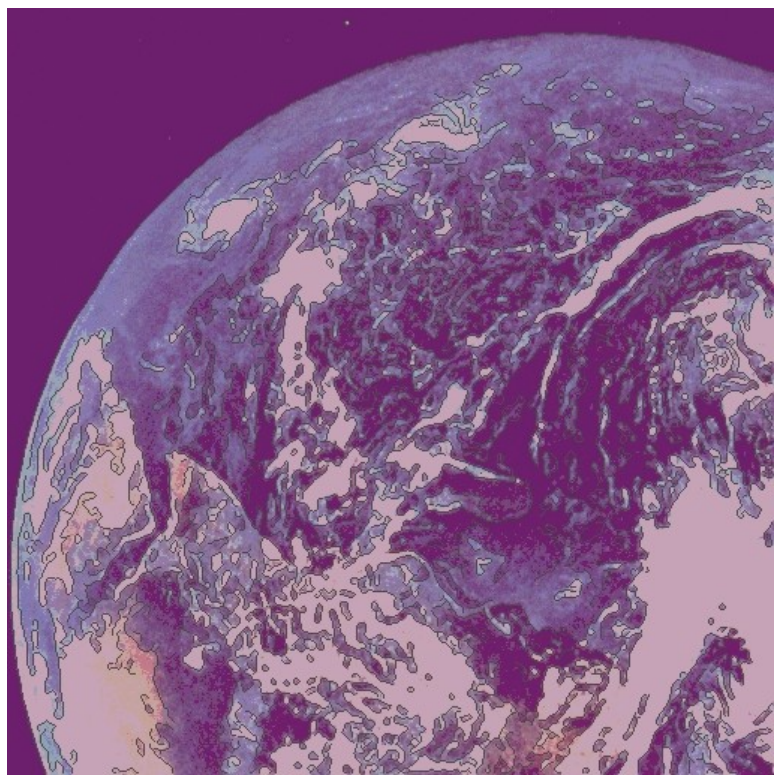

Fig.6. (g)

Fig.5. Results:(a) Original Earth Image.

(b,c,d) Segmentation in space domain at threshold level 3,5 and 7.

(e,f,g) Segmentation by proposed algorithm at threshold level 3,5 and 7.

In Fig. 7, histograms of Approximation, Horizontal, Vertical and Diagonal coefficients of R, G and B components of original and segmented Earth image in wavelet domain is depicted.

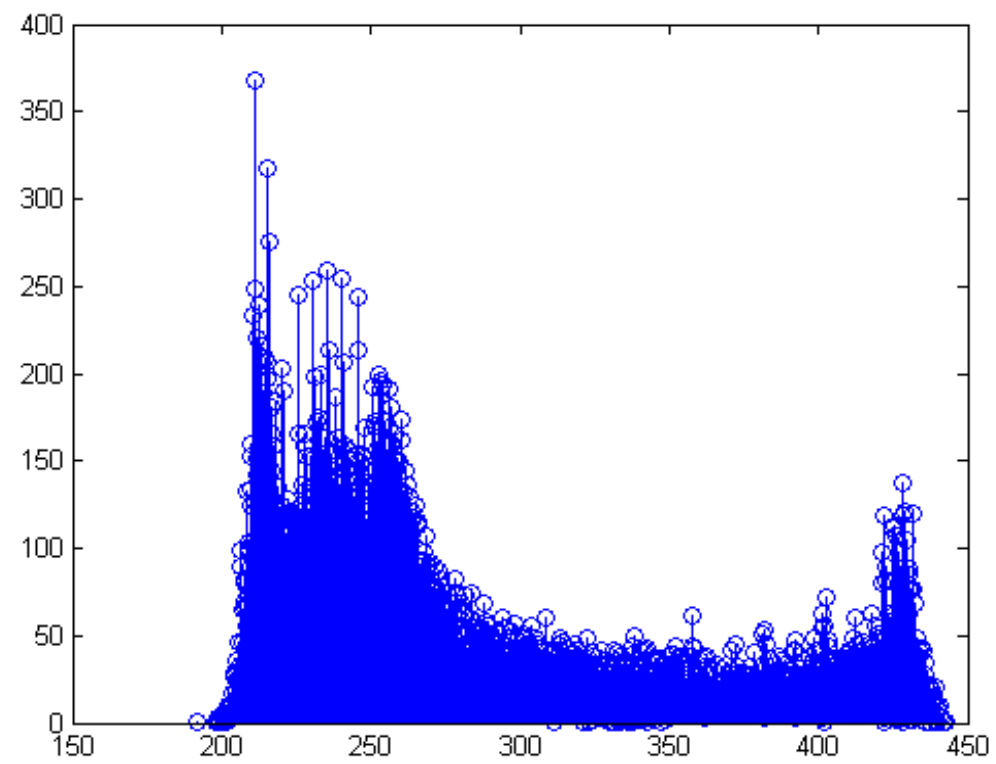

Fig.7. (1)

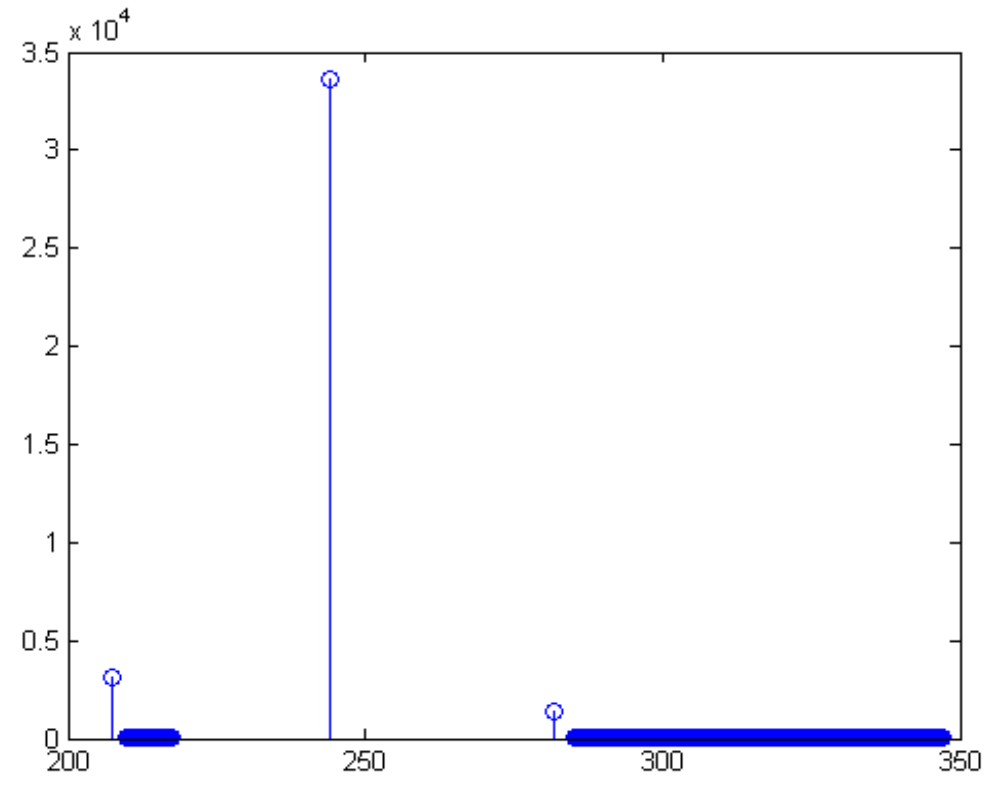

Fig.7. (2)

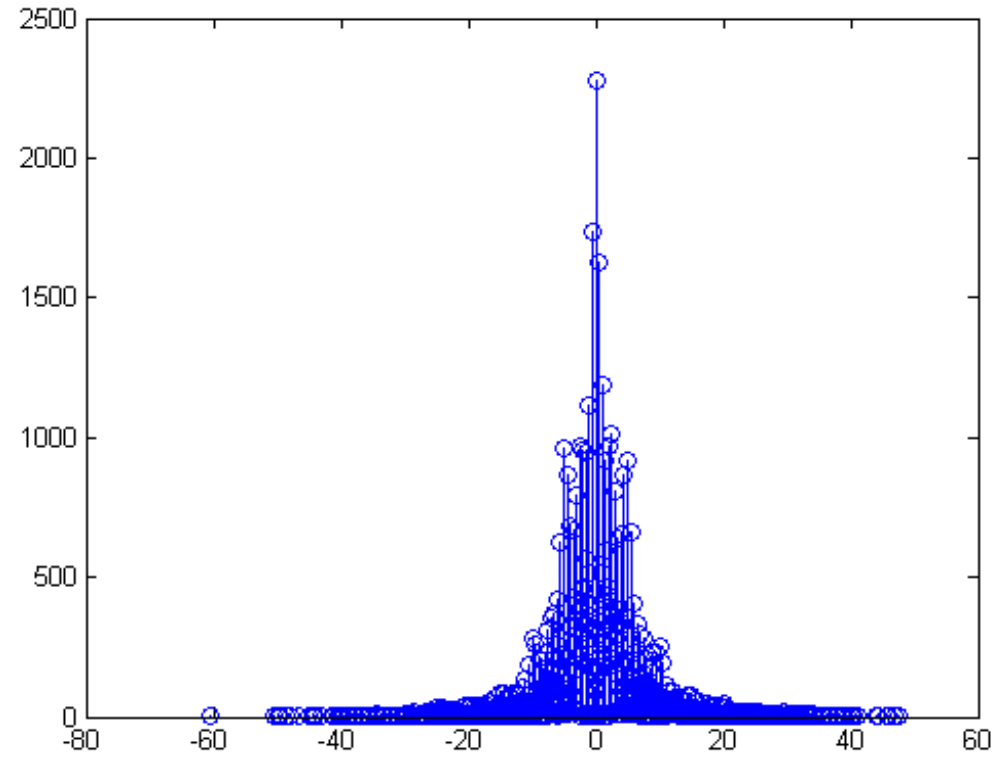

Fig.7. (3)

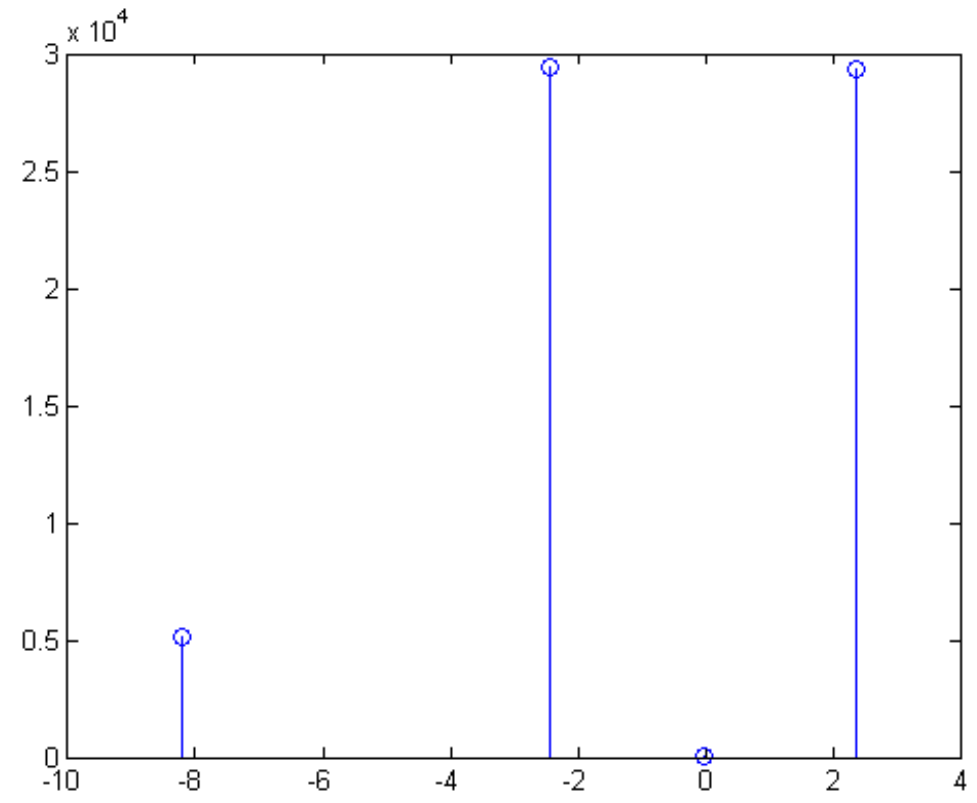

Fig.7. (4)

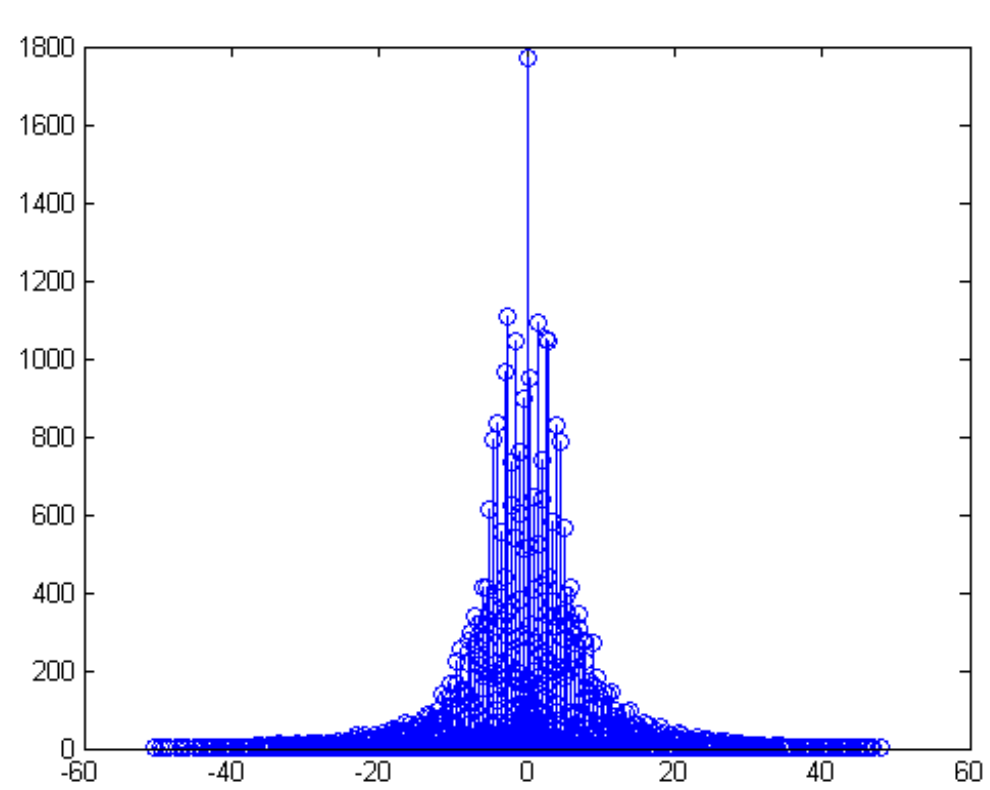

Fig.7. (5)

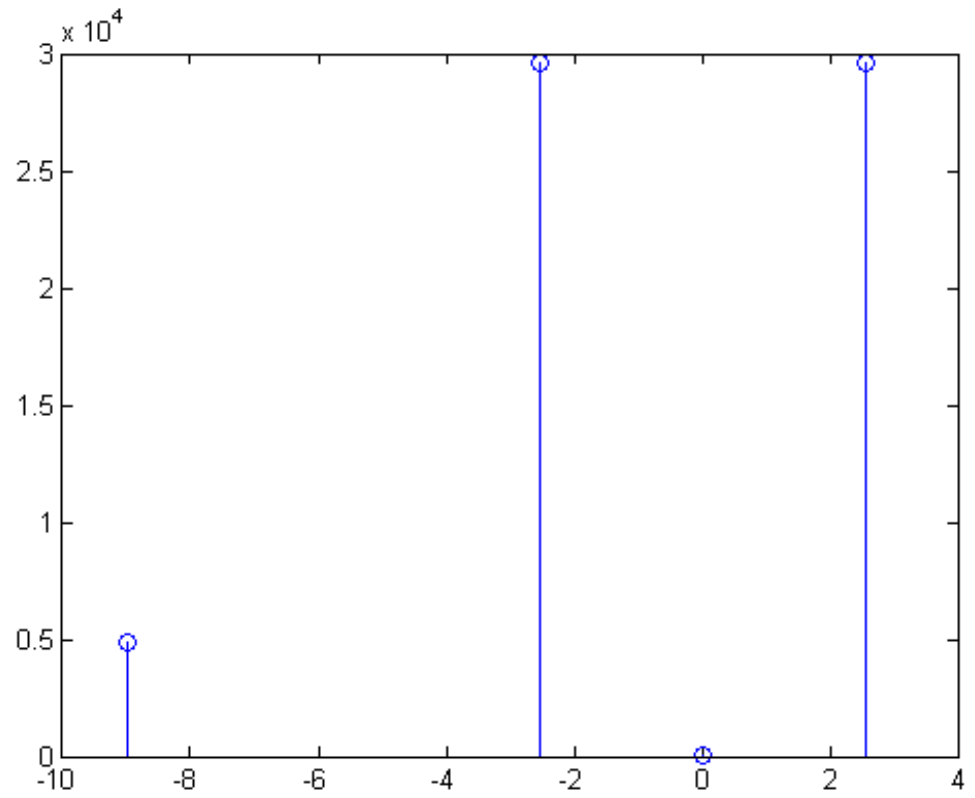

Fig.7. (6)

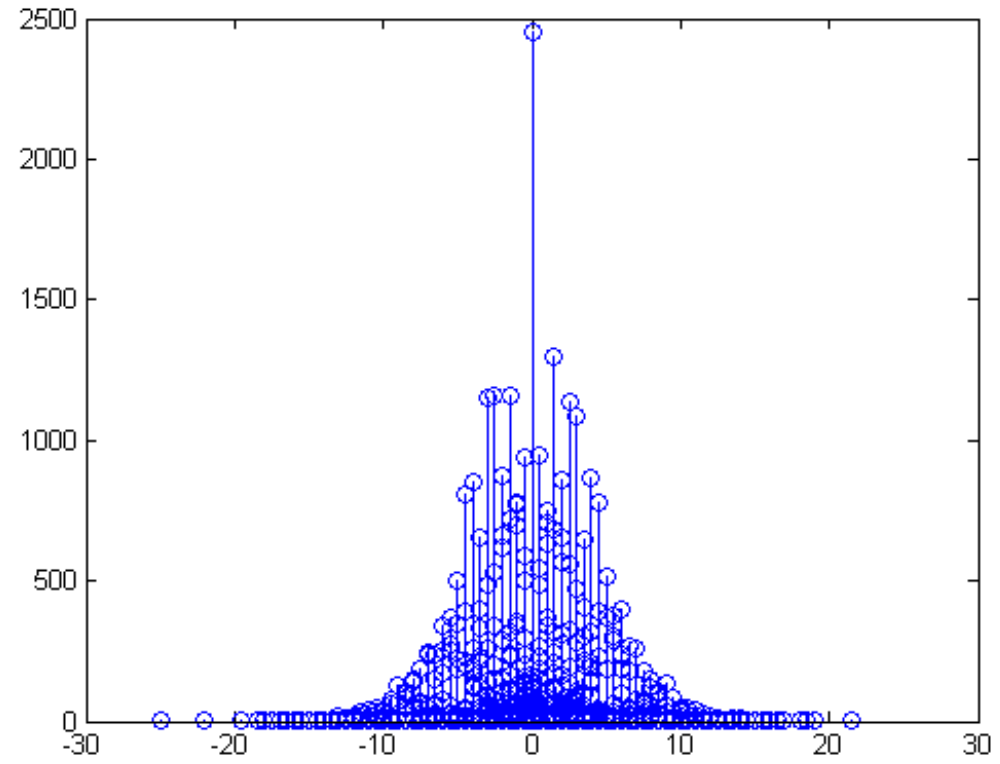

Fig.7. (7)

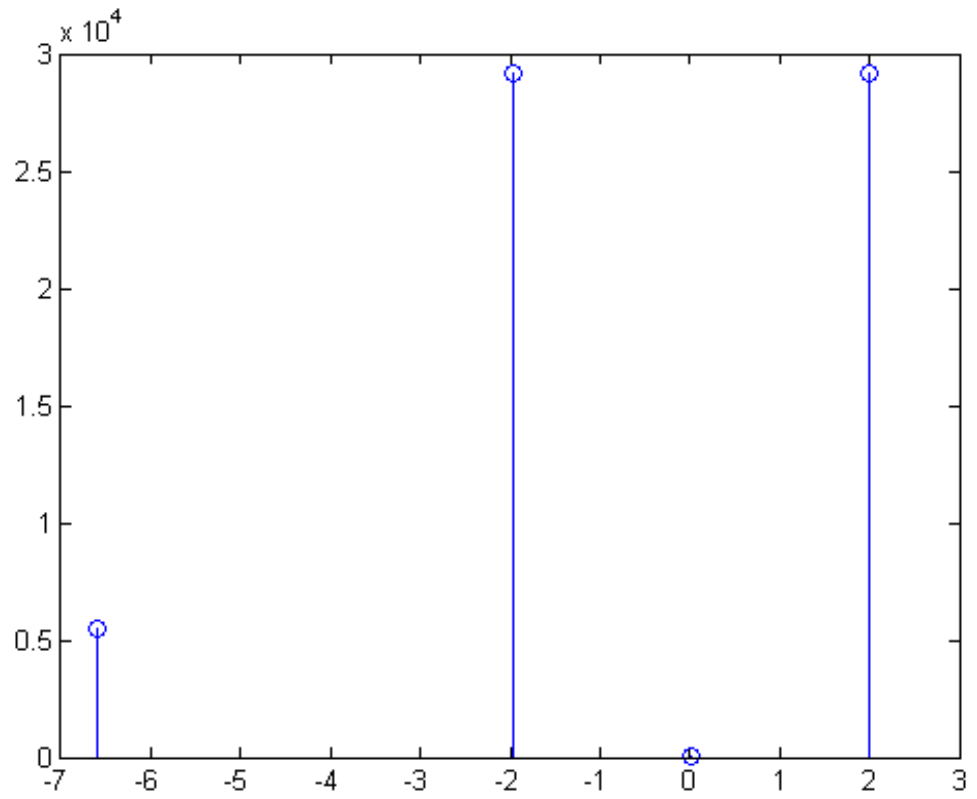

Fig.7. (8)

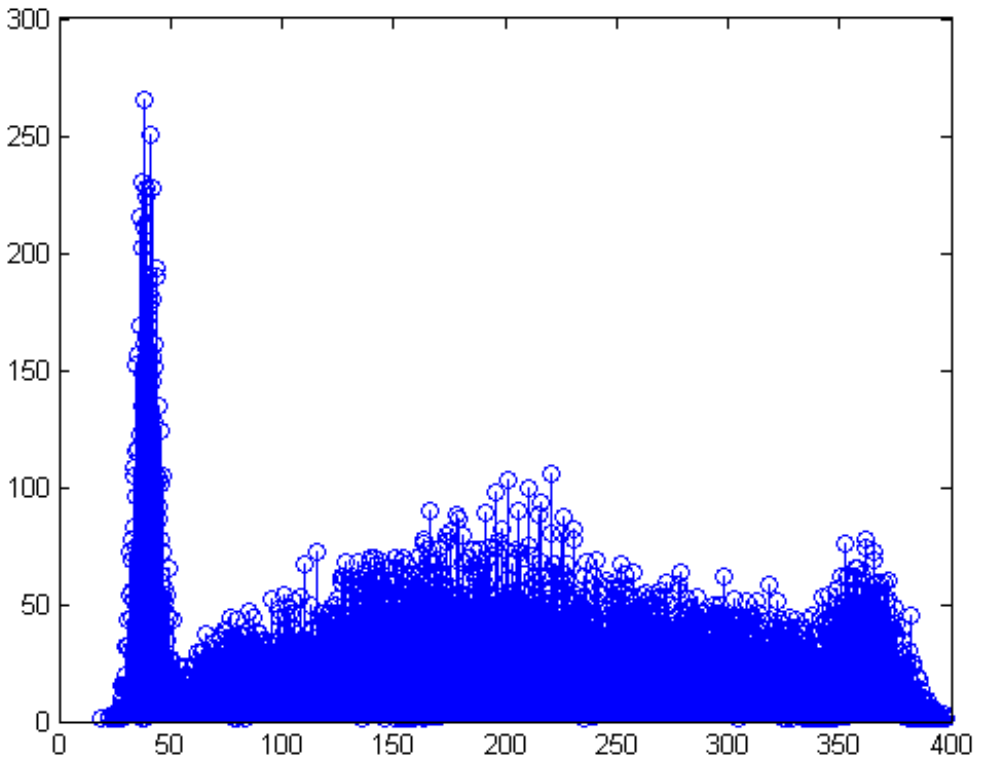

Fig.7. (9)

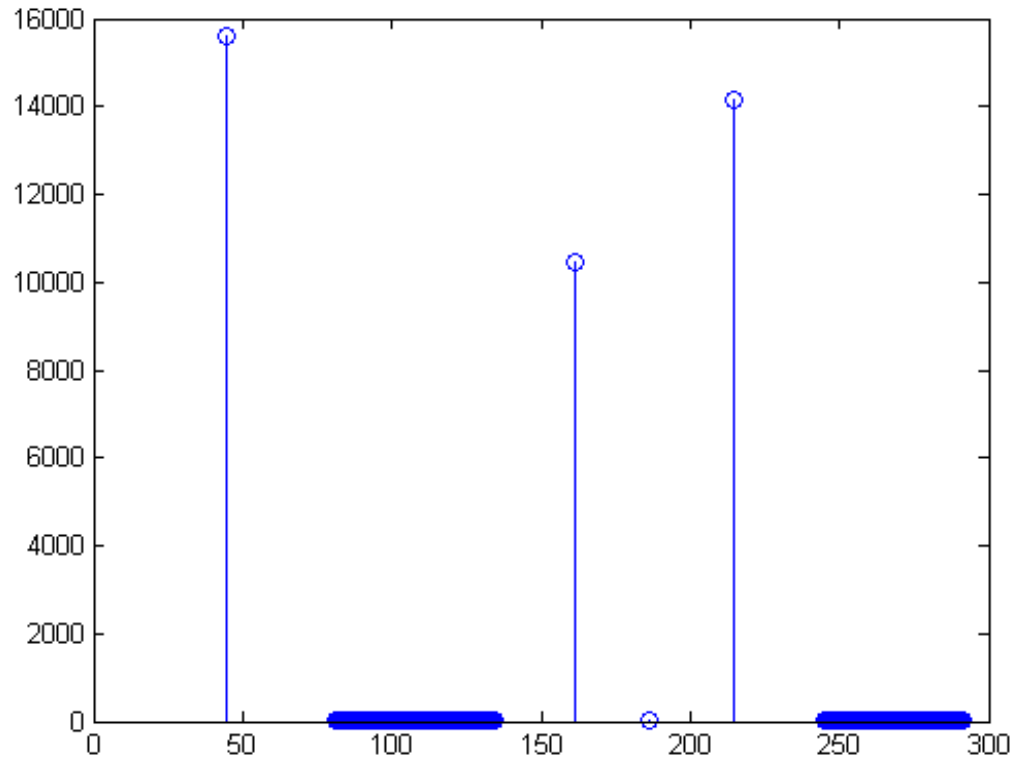

Fig.7. (10)

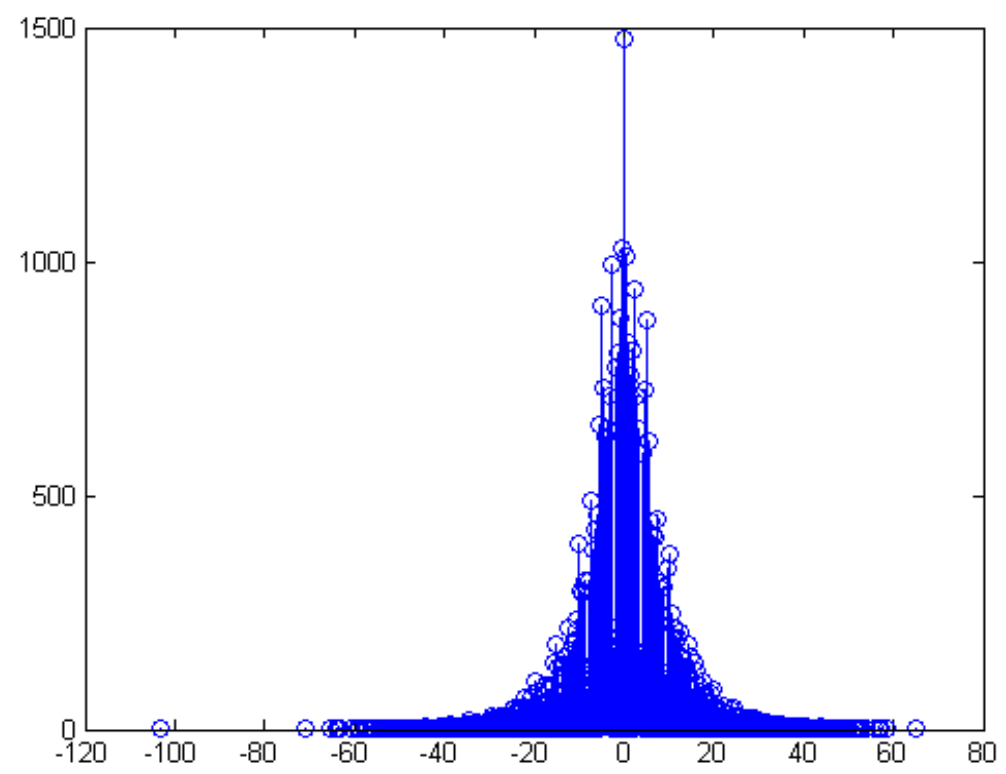

Fig.7. (11)

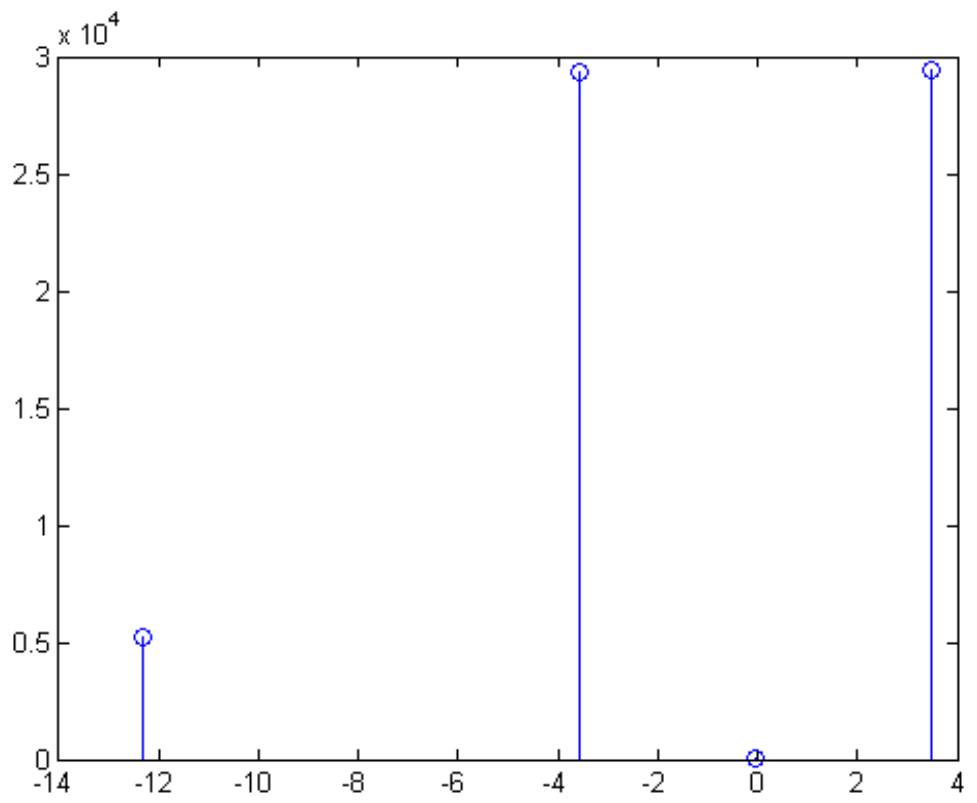

Fig.7. (12)

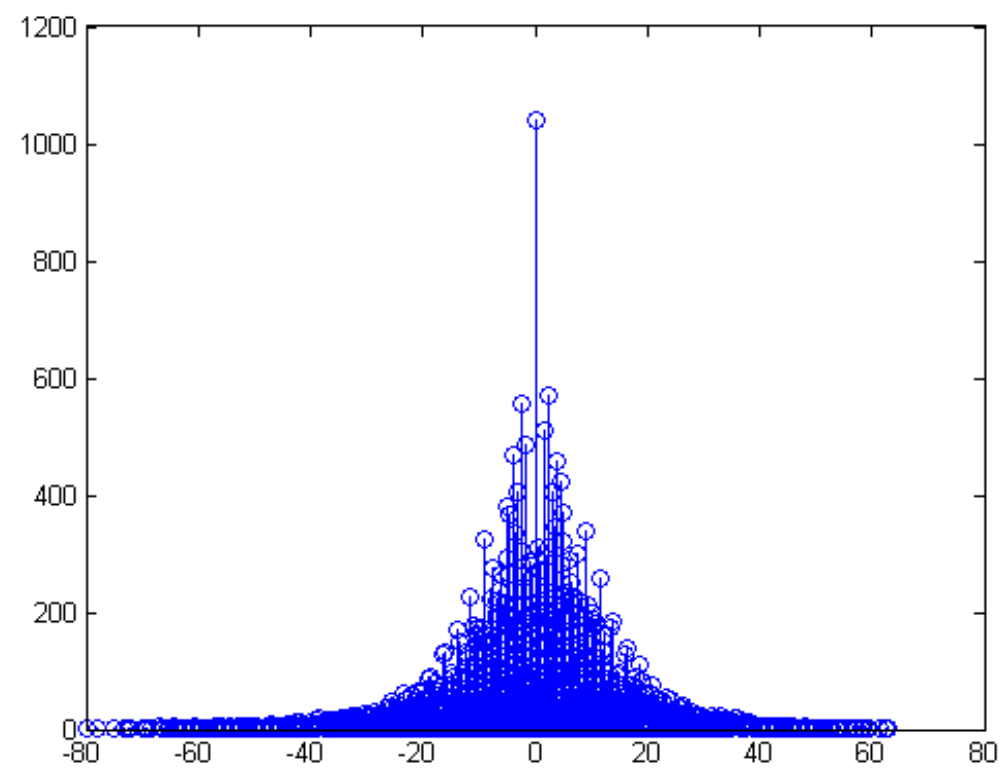

Fig.7. (13)

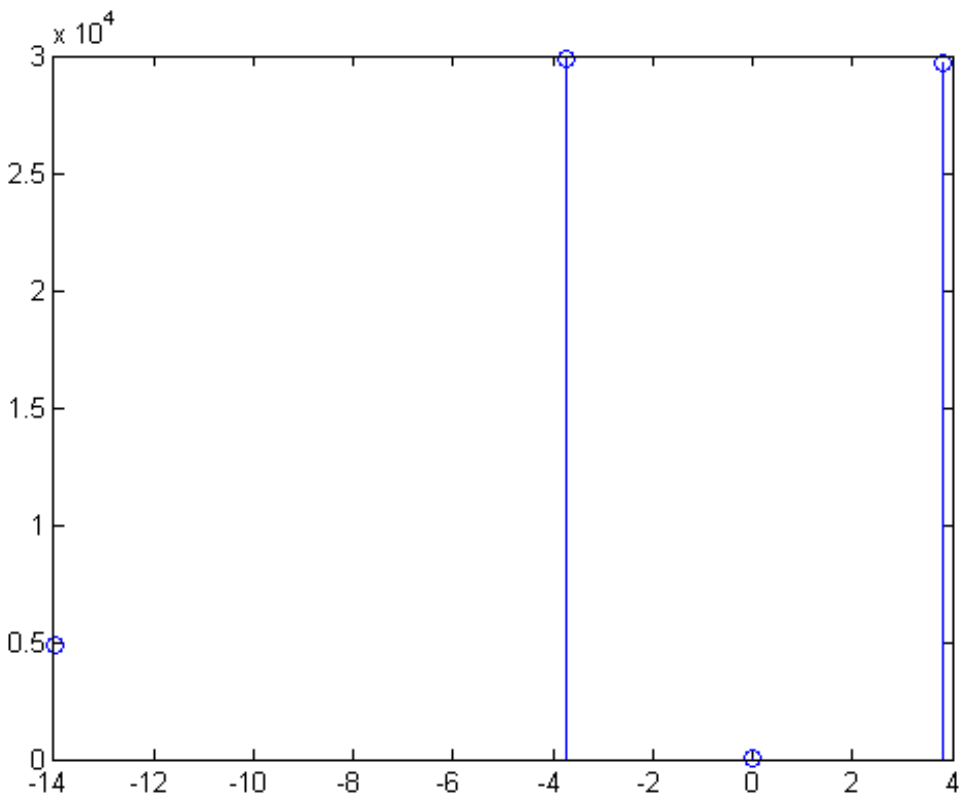

Fig.7. (14)

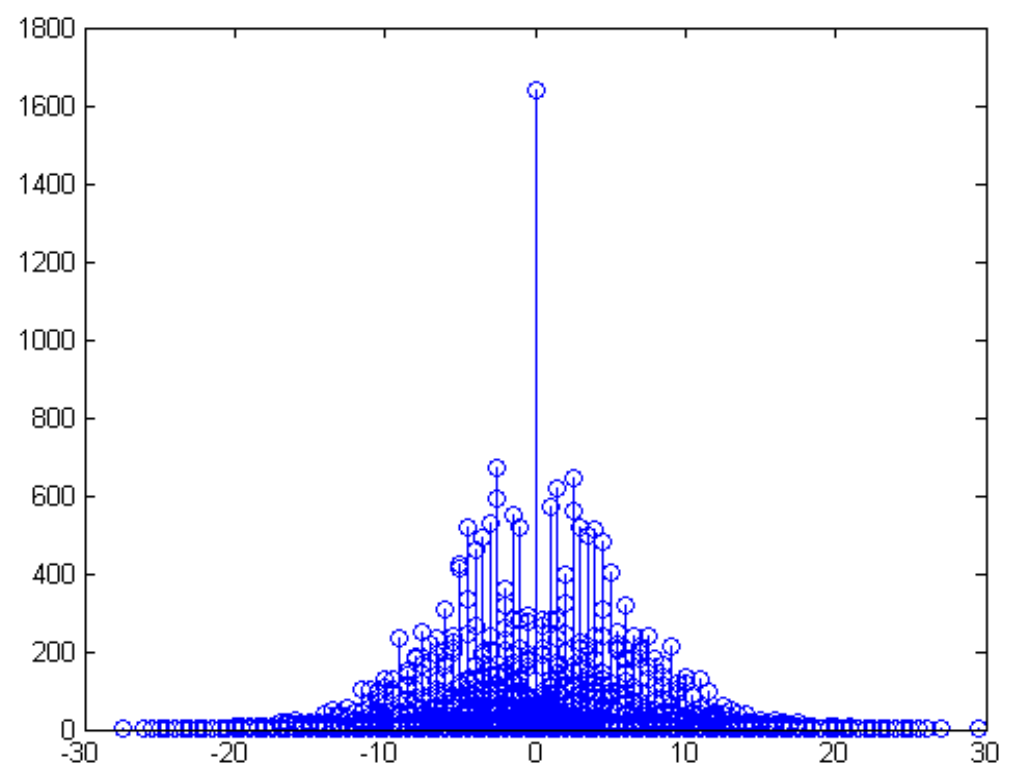

Fig.7. (15)

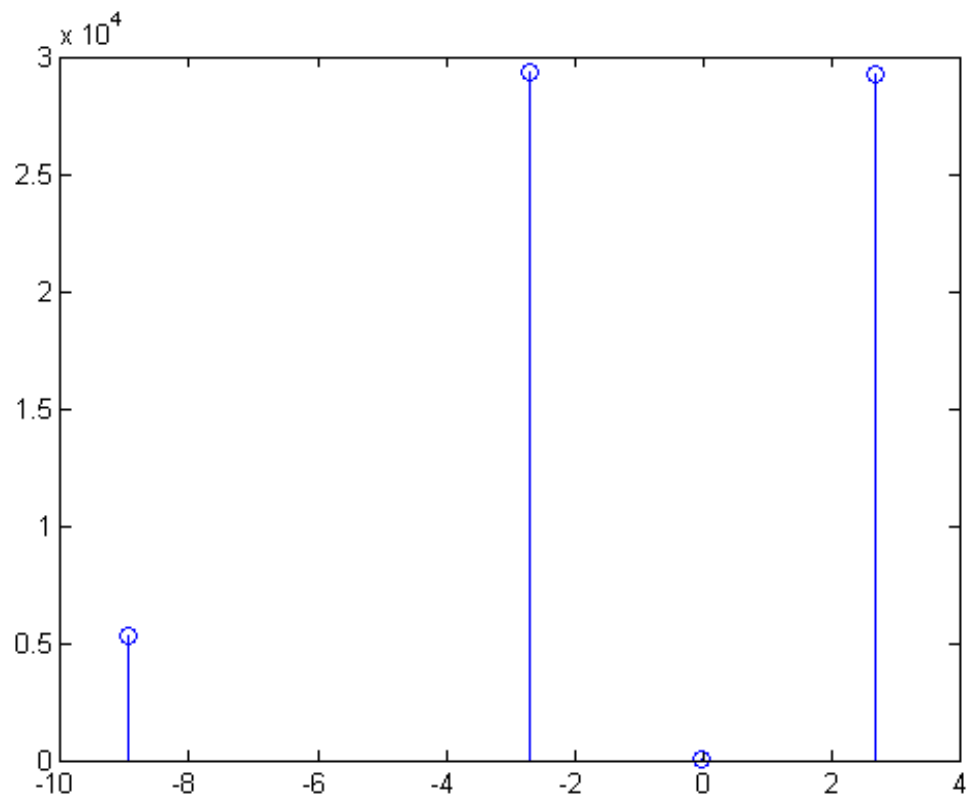

Fig.7. (16)

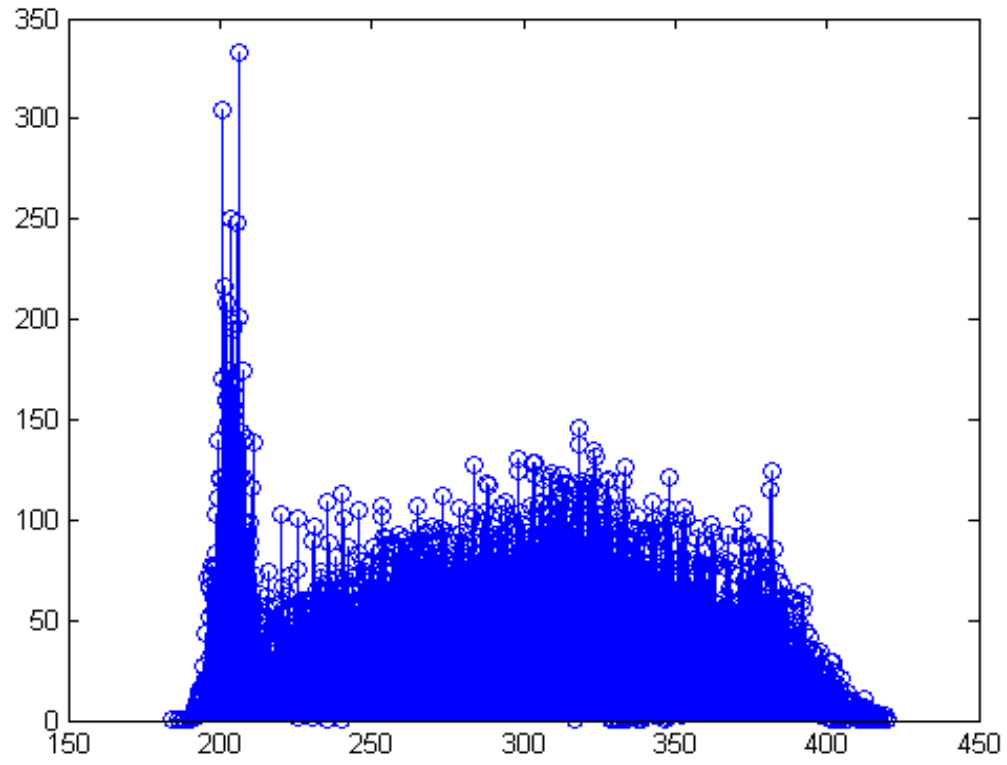

Fig.7. (17)

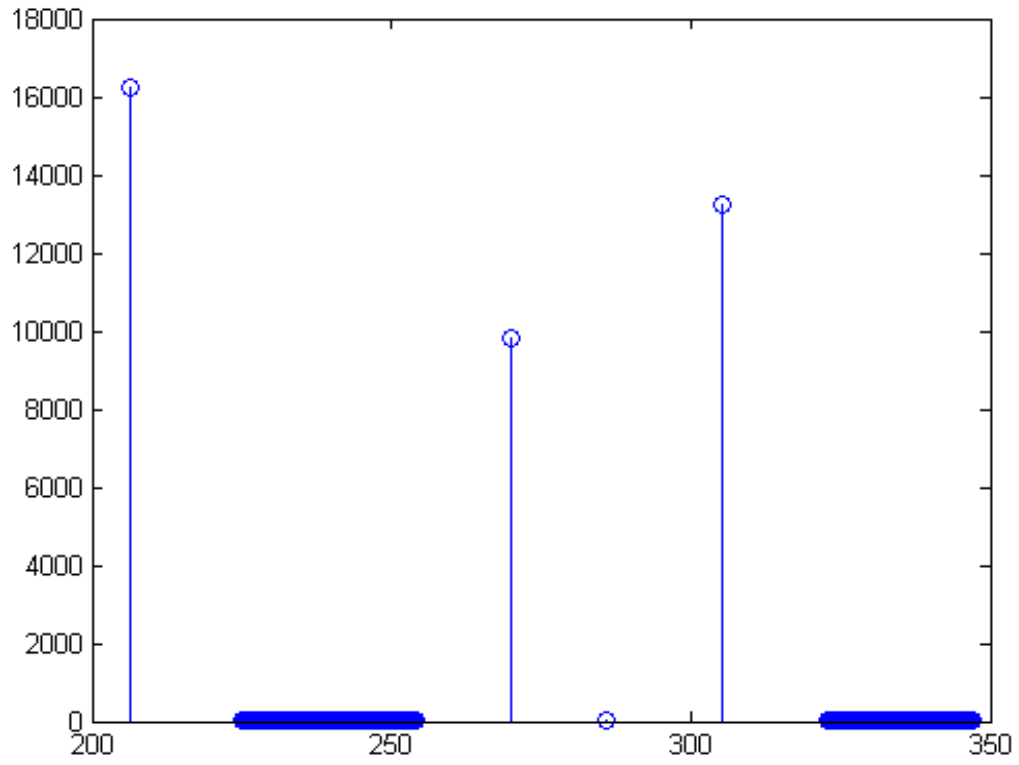

Fig.7. (18)

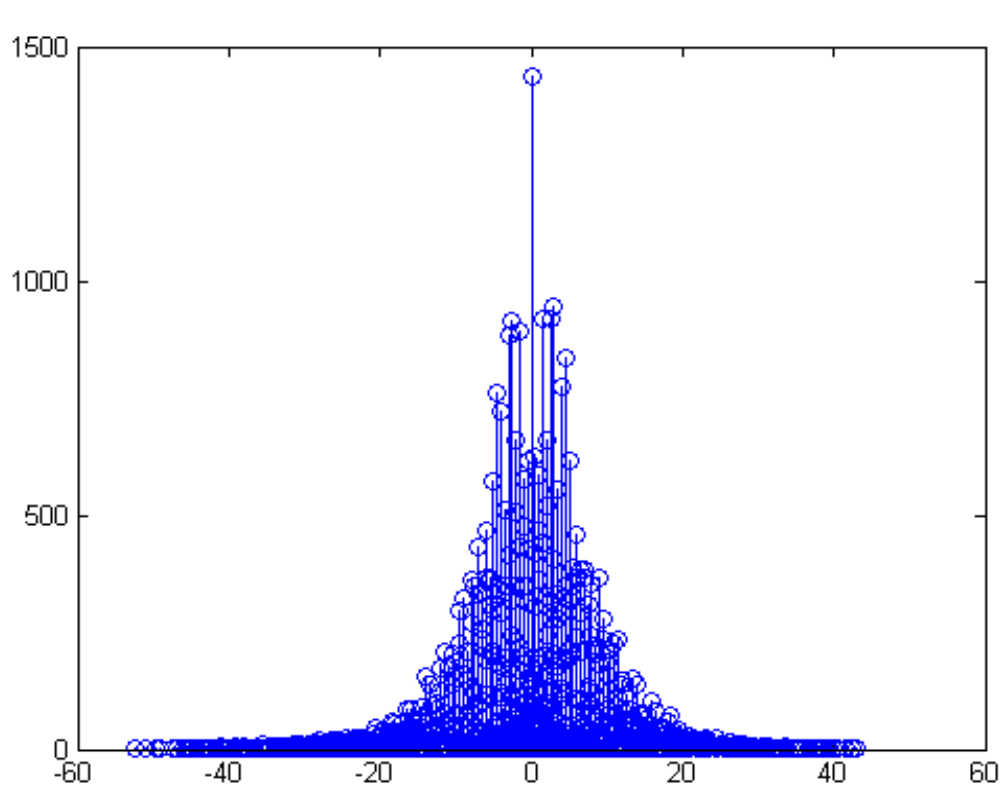

Fig.7. (19)

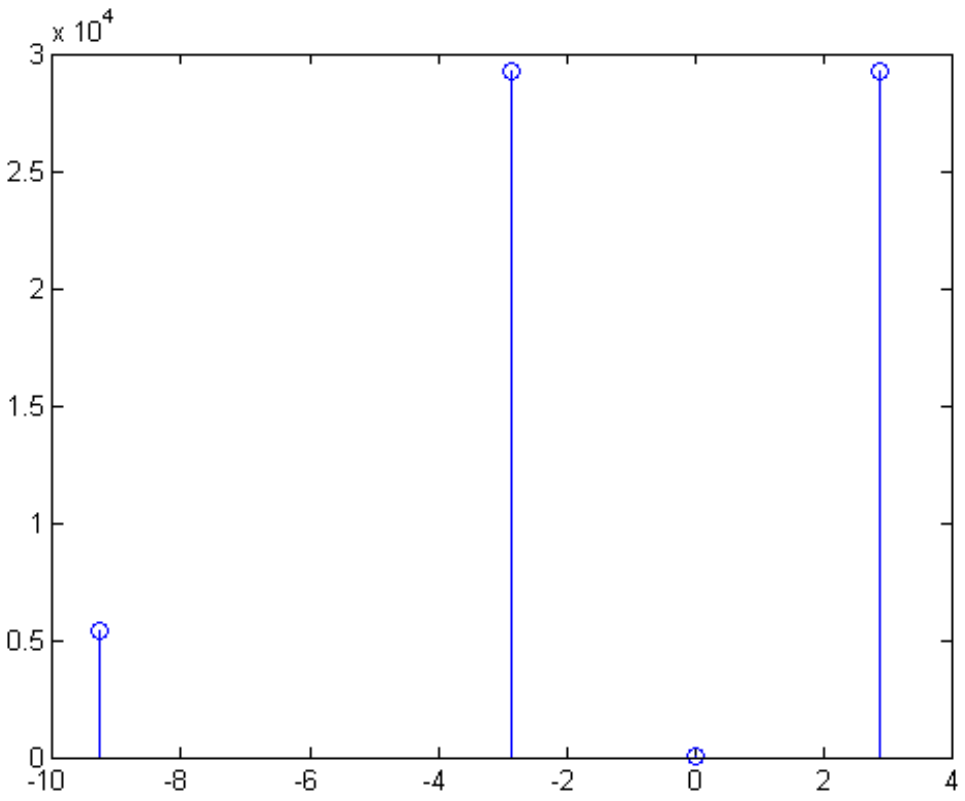

Fig.7. (20)

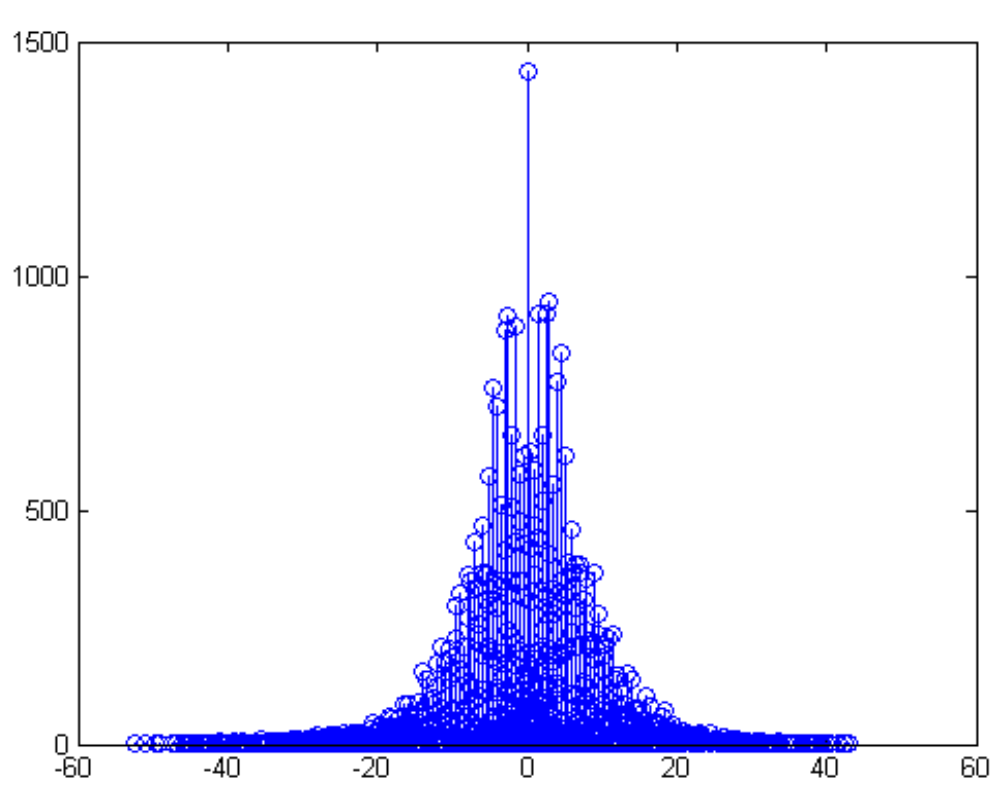

Fig.7. (21)

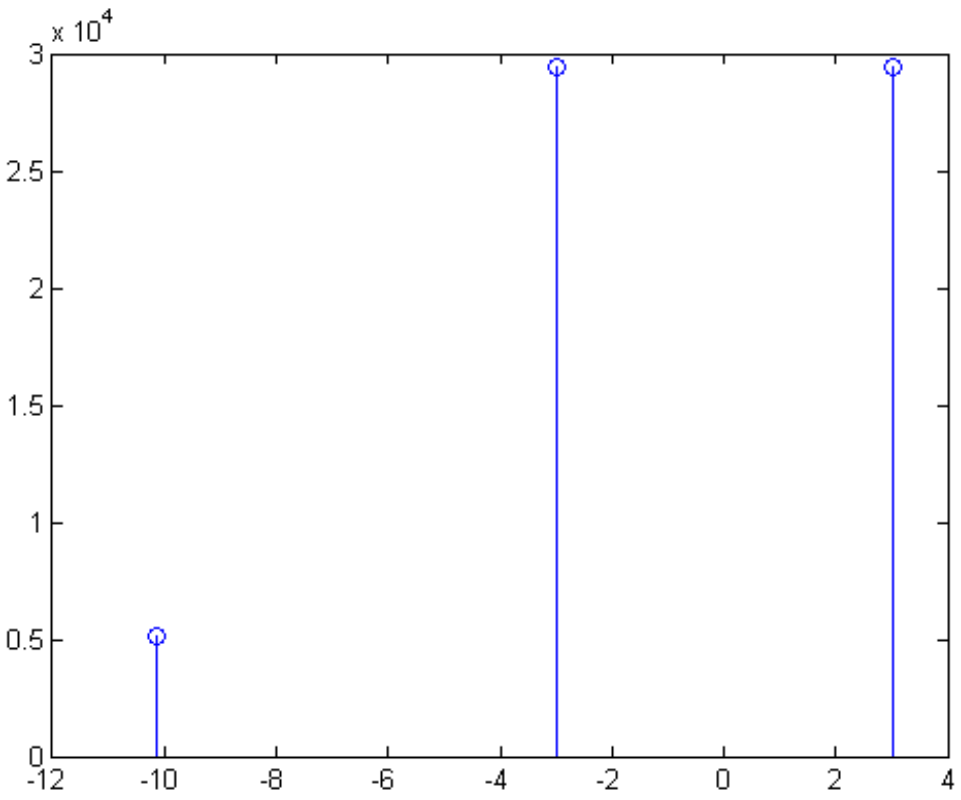

Fig.7. (22)

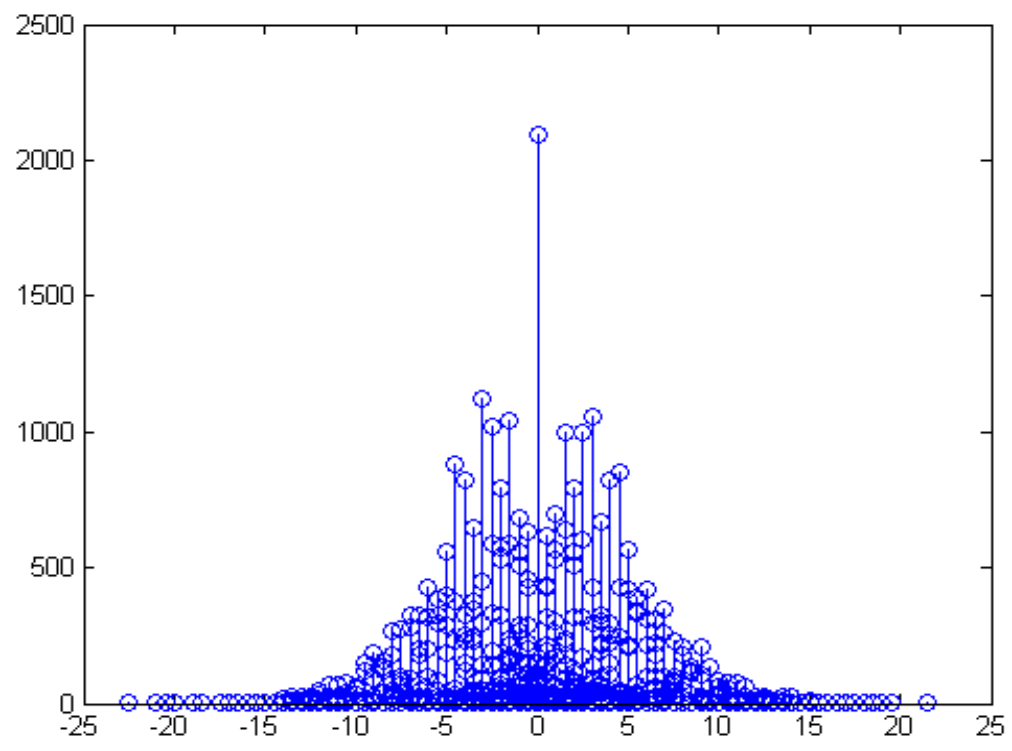

Fig.7. (23)

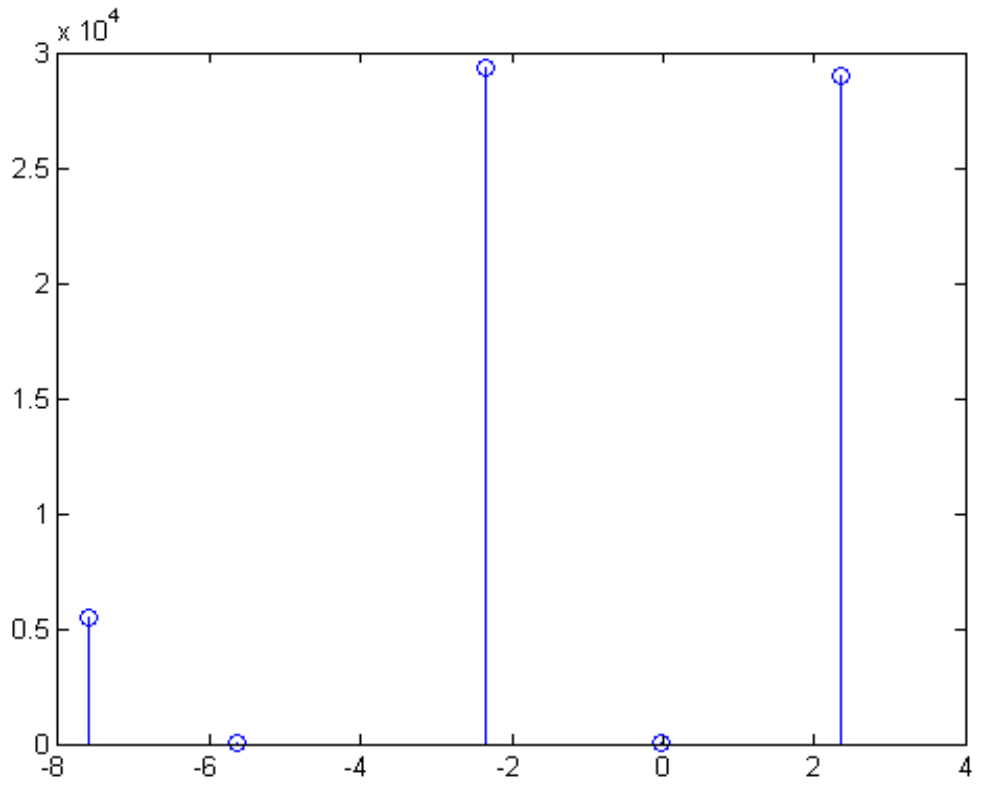

Fig.7. (24)

Fig.7. Results:(1,9,17) Histogram of Approximation coefficients of R, G and B components respectively in wavelet domain of Earth image.

(2,10,18) Histogram of Approximation coefficients of R, G and B components respectively in wavelet domain of Earth image thresholded at level 3.

(3,11,19) Histogram of Horizontal coefficients of R, G and B components respectively in wavelet domain of Earth image.

(4,12,20) Histogram of Horizontal coefficients of R, G and B components respectively in wavelet domain of Earth image thresholded at level 3.

(5,13,21) Histogram of Vertical coefficients of R, G and B components respectively in wavelet domain of Earth image.

(6,14,22) Histogram of Vertical coefficients of R, G and B

components respectively in wavelet domain of Earth image thresholded at level 3.

(7,15,23) Histogram of Diagonal coefficients of R, G and B components respectively in wavelet domain of Earth image.

(8,16,24) Histogram of Diagonal coefficients of R, G and B components respectively in wavelet domain of Earth image thresholded at level 3.

Fig. 8 shows the histogram of segmented R, G and B components of Aerial image at thresholding levels 3, 5 and 7.

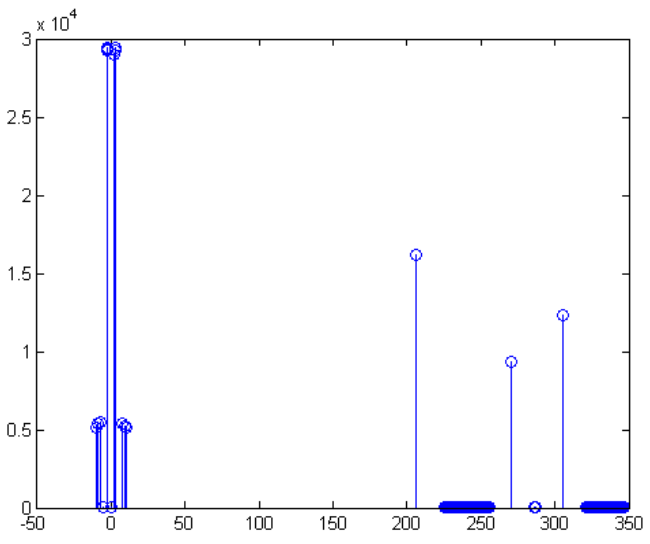

Fig.8. (a)

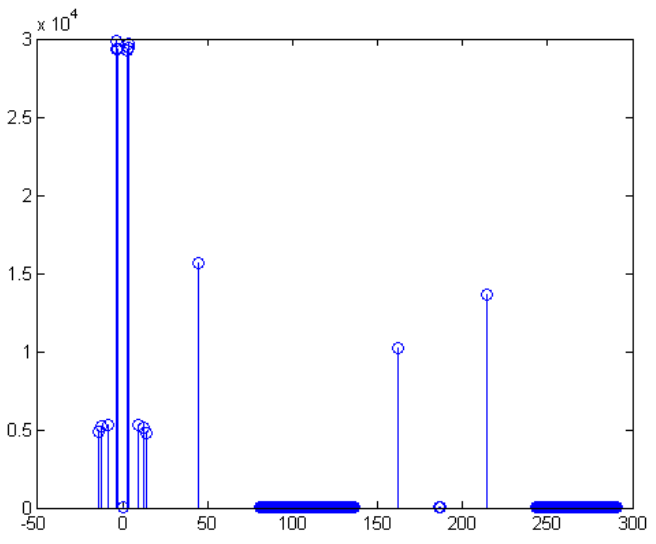

Fig.8. (b)

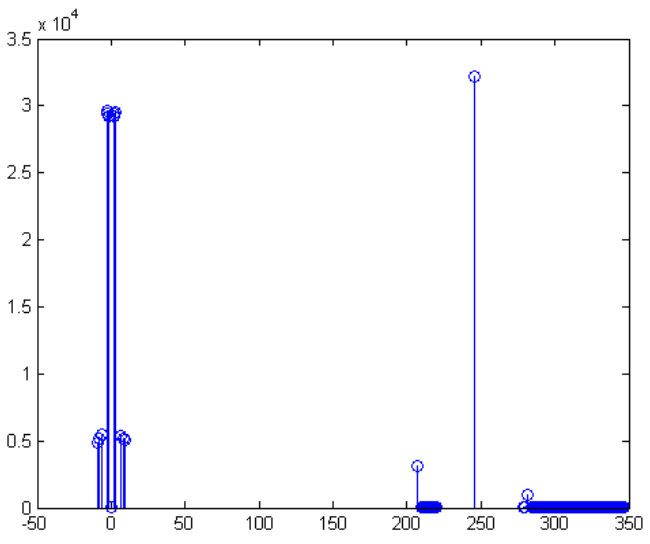

Fig.8. (c)

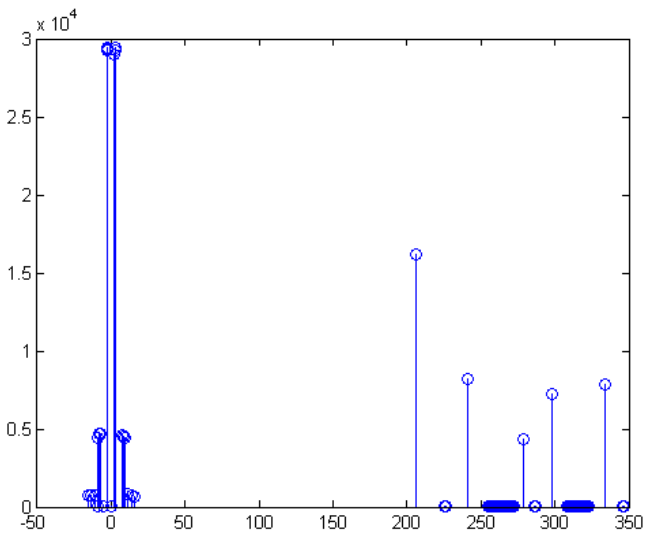

Fig.8. (d)

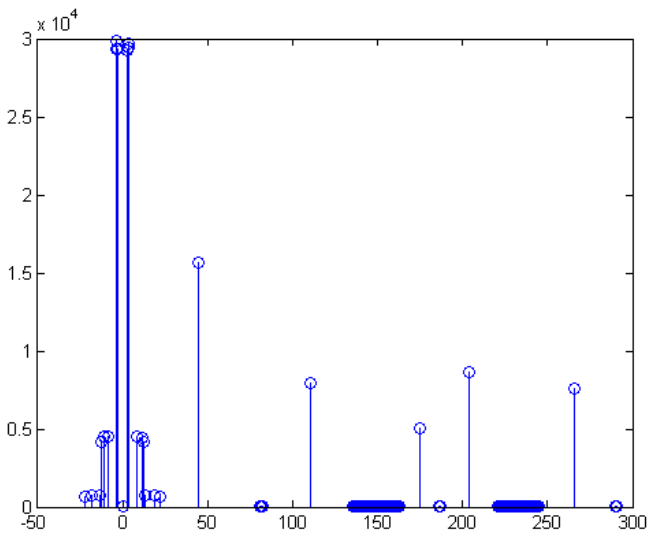

Fig.8. (e)

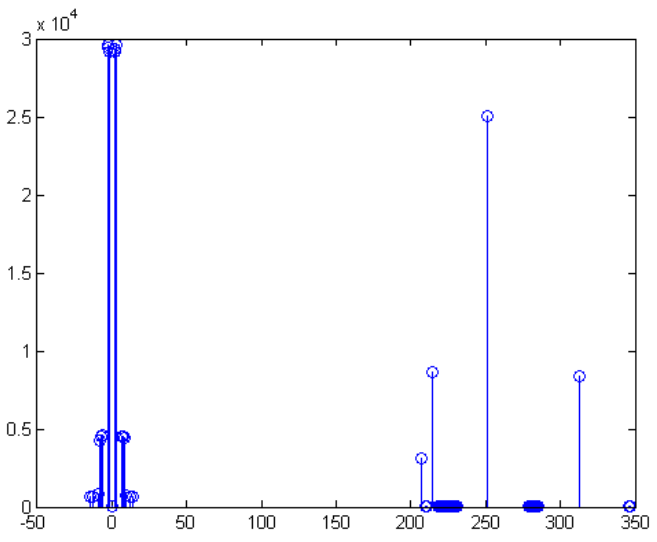

Fig.8. (f)

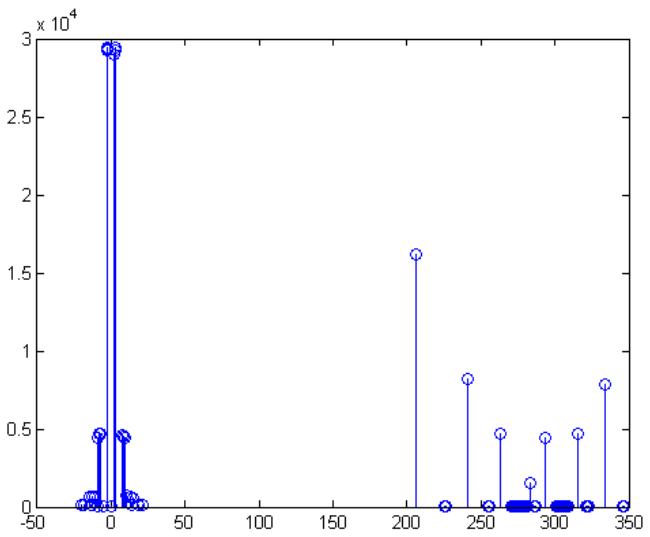

Fig.8. (g)

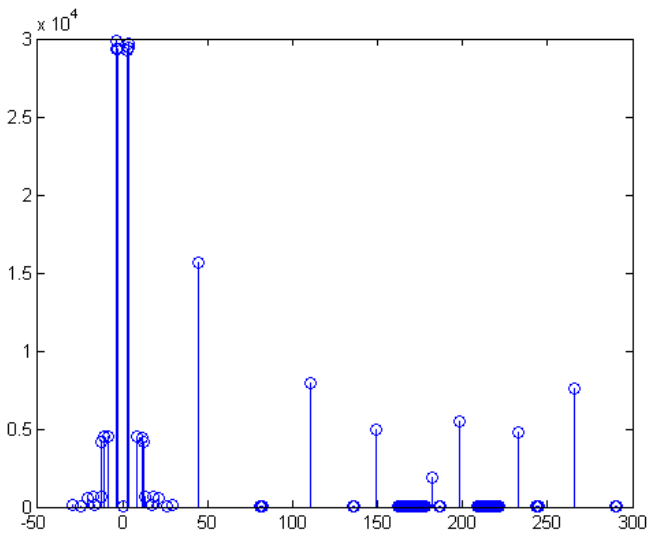

Fig.8. (h)

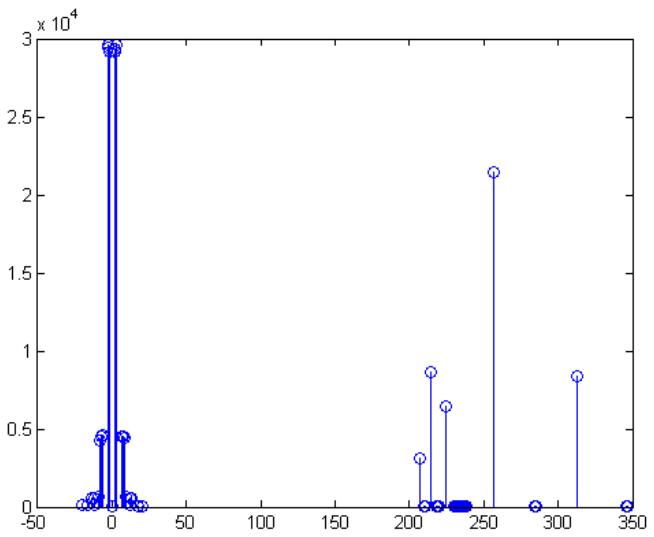

Fig.8. (i)

Fig.8. Results:(a,b,c) Histogram of B, G and R components thresholded in wavelet domain at level 3 of Earth image.

(d,e,f) Histogram of B, G and R components thresholded in wavelet domain at level 5 of Earth image.

(g,h,i) Histogram of B, G and R components thresholded in wavelet domain at level 7 of Earth image.

*Table 4: Comparison of SSIM of Earth image (512 x512, 768.1 kB) between image segmentation in space domain and by proposed algorithm.*

| Threshold Level | SSIM in Space Domain | SSIM in Wavelet Domain |
|---|---|---|
| 3 | 0.9685 | 0.9806 |
| 5 | 0.9672 | 0.9796 |
| 7 | 0.9674 | 0.9797 |

*Table 5: Comparison of PSNR of Earth image (512 x512, 768.1 kB) between image segmentation in space domain and by proposed algorithm.*

| Threshold Level | PSNR in Space Domain (dB) | PSNR in Wavelet Domain (dB) |
|---|---|---|
| 3 | 24.1114 | 25.9095 |
| 5 | 23.8892 | 26.2669 |
| 7 | 23.8944 | 26.4056 |

*Table 6: Comparison of Time Complexity of Earth image (512 x512, 768.1 kB) between image segmentation in space domain and by proposed algorithm.*

| Threshold Level | Time Complexity in Space Domain (sec) | Time Complexity in Wavelet Domain (sec) |
|---|---|---|
| 3 | 3.0264 | 1.6692 |
| 5 | 3.6036 | 1.7628 |
| 7 | 3.9312 | 1.8564 |

The results of Table 4, 5 and 6 are plotted in the fig. 9. The red color graph represents the result of space domain algorithm while the black color represents the results of proposed algorithm.

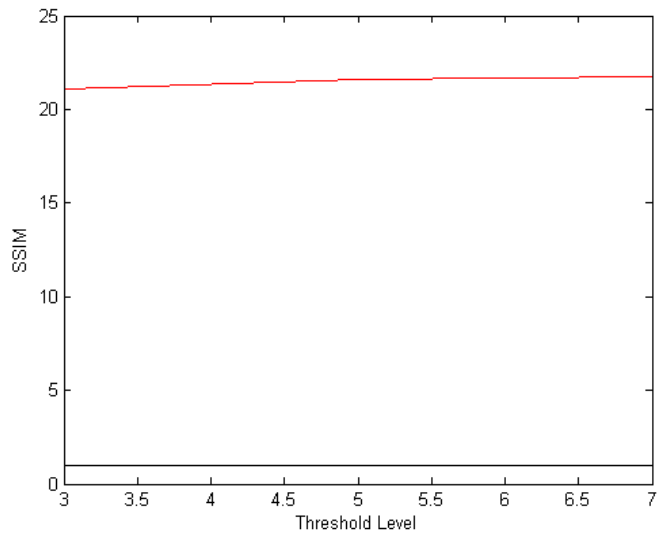

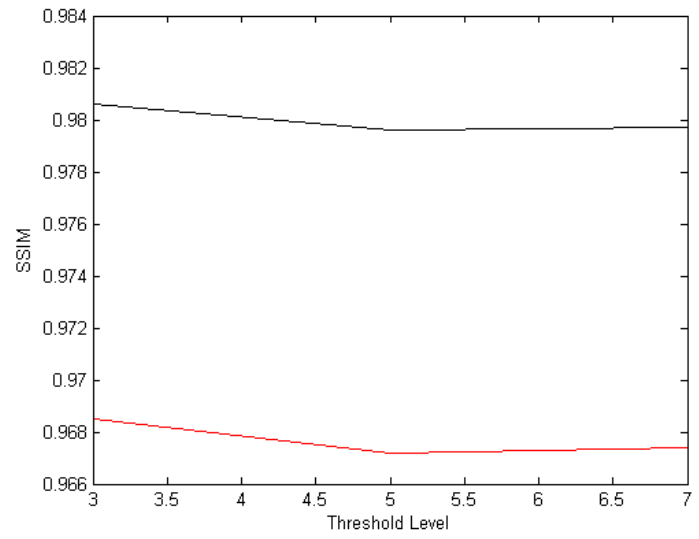

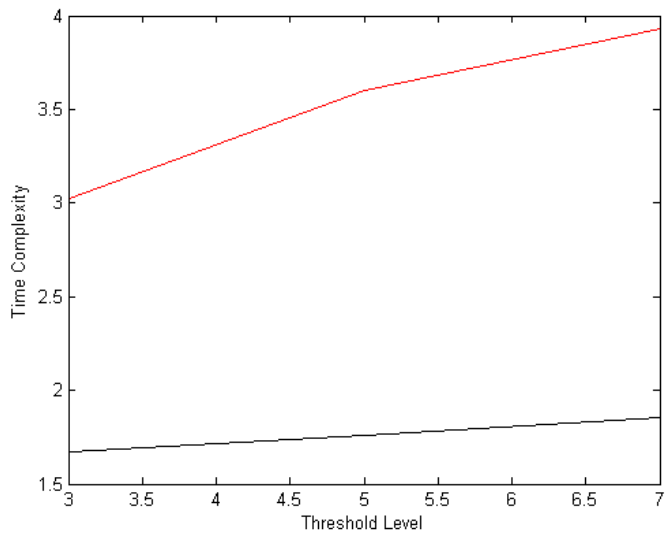

Fig.9. (a) Fig.9. (b) Fig.9. (c)

Fig.9. Results: (a) Comparison of PSNR of Earth image between image segmentation in space domain and by proposed algorithm.

(b) Comparison of SSIM of Earth image between image segmentation in space domain and by proposed algorithm.

(c) Comparison of Time Complexity of Earth image between image segmentation in space domain and by proposed algorithm.

Table 7: Comparison of PSNR, SSIM and Time Complexity of various test images between image segmentation in space domain and by proposed algorithm.

| Image Name | Threshold Level | SSIM in Space Domain | SSIM in Wavelet Domain | PSNR in Space Domain (dB) | PSNR in Wavelet Domain (dB) | Time Complexity in Space Domain (sec) | Time Complexity in Wavelet Domain (sec) |
|---|---|---|---|---|---|---|---|
| House | 3 | 0.9870 | 0.9875 | 22.5739 | 23.3216 | 0.8268 | 0.3276 |
| | 5 | 0.9846 | 0.9870 | 22.4972 | 23.6323 | 0.8892 | 0.3900 |
| 256x256 | 7 | 0.9846 | 0.9872 | 22.5652 | 23.7875 | 0.9672 | 0.3744 |
| | | | | | | | |
| Lenna | 3 | 0.9783 | 0.9863 | 21.8314 | 23.8069 | 2.8392 | 1.7160 |
| | 5 | 0.9776 | 0.9866 | 21.6453 | 24.1578 | 3.4164 | 1.7472 |
| 512x512 | 7 | 0.9776 | 0.9874 | 21.6396 | 24.3705 | 3.6972 | 1.9188 |
| | | | | | | | |
| Pepper | 3 | 0.9758 | 0.9823 | 19.6385 | 21.6214 | 2.8236 | 1.6224 |
| | 5 | 0.9755 | 0.9859 | 19.5087 | 22.4168 | 3.3852 | 1.7472 |
| 512x512 | 7 | 0.9754 | 0.9866 | 19.5033 | 22.5544 | 3.6504 | 1.9032 |
| | | | | | | | |
| Baboon | 3 | 0.9667 | 0.9742 | 20.4878 | 21.0555 | 3.0108 | 1.6848 |
| | 5 | 0.9646 | 0.9756 | 20.3132 | 21.6169 | 3.5568 | 1.7628 |
| 512x512 | 7 | 0.9645 | 0.9759 | 20.3057 | 21.7436 | 3.9312 | 1.8564 |

The results of Table 7 are plotted in the fig. 10. The red color graph represents the result of space domain algorithm while the black color represents the results of proposed algorithm.

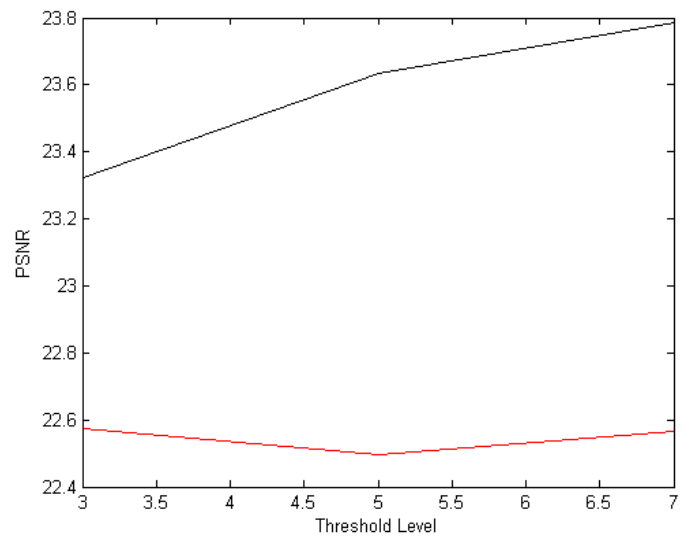

Fig.10. (a)

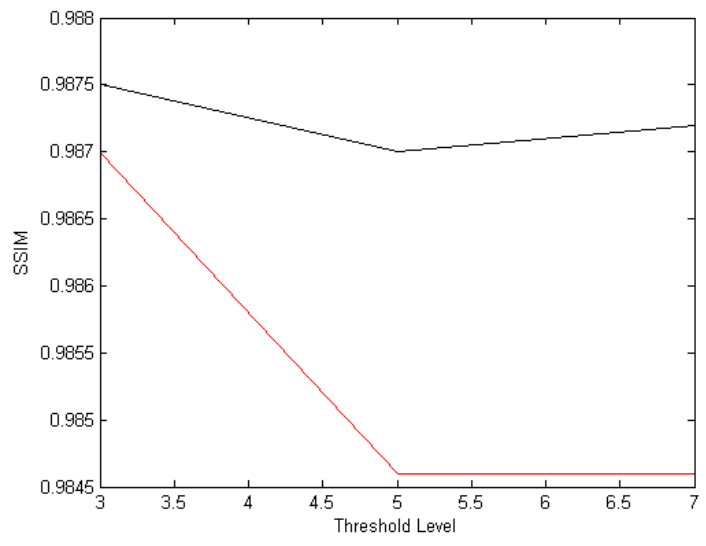

Fig.10. (b)

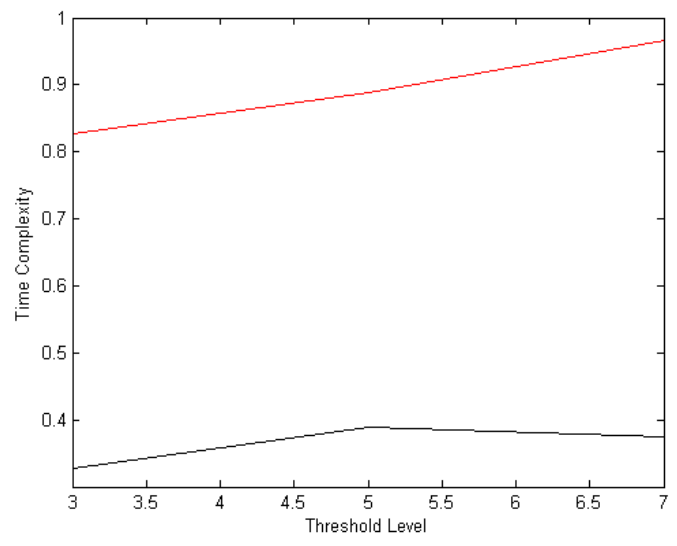

Fig.10. (c)

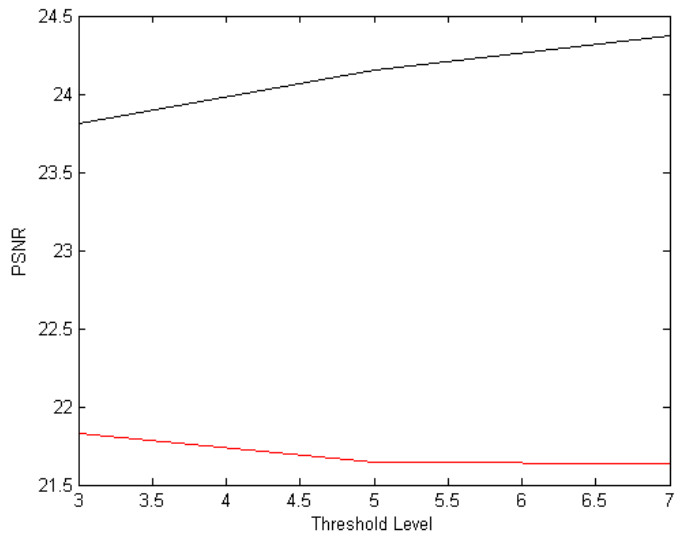

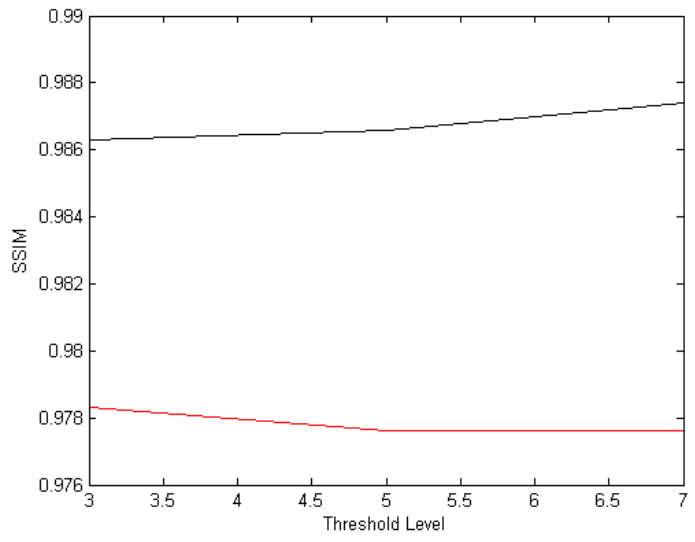

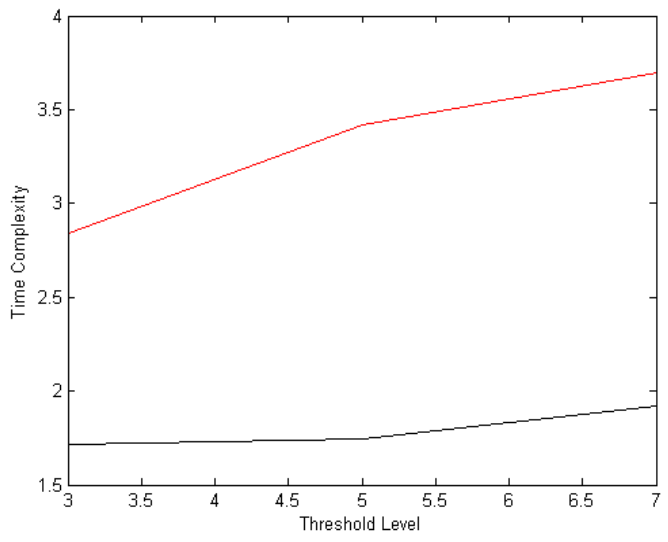

Fig.10. (d) Fig.10. (e) Fig.10. (f)

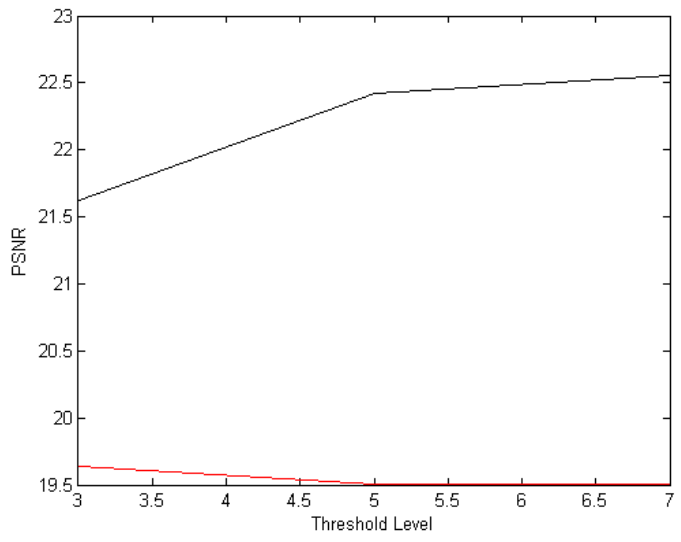

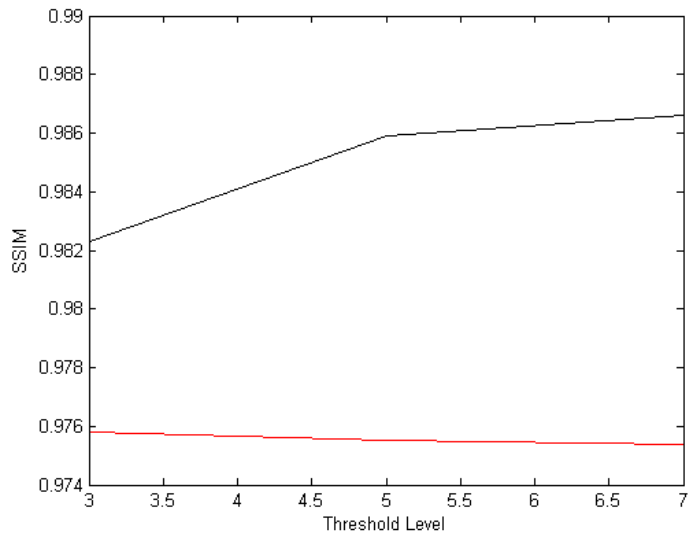

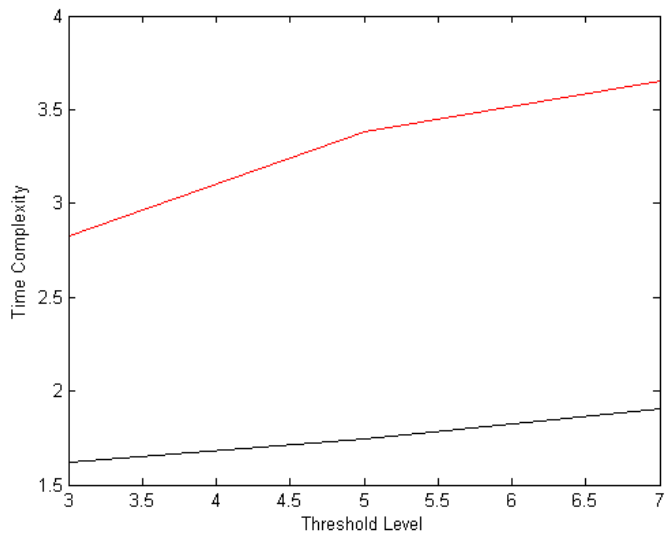

Fig.10. (g) Fig.10. (h) Fig.10. (i)

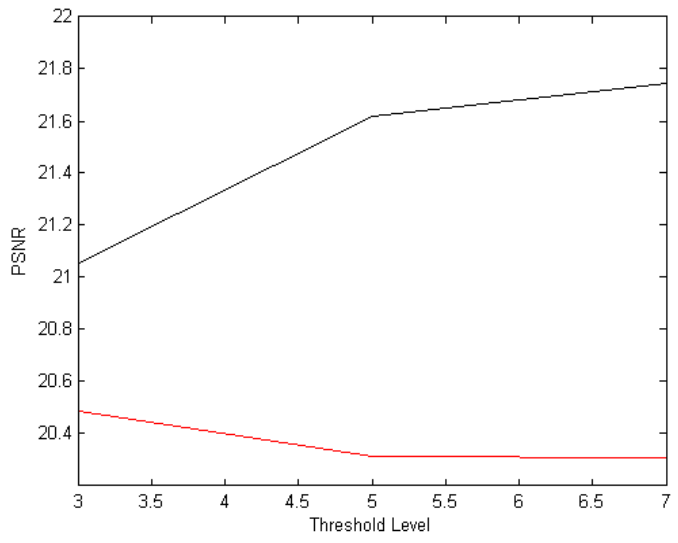

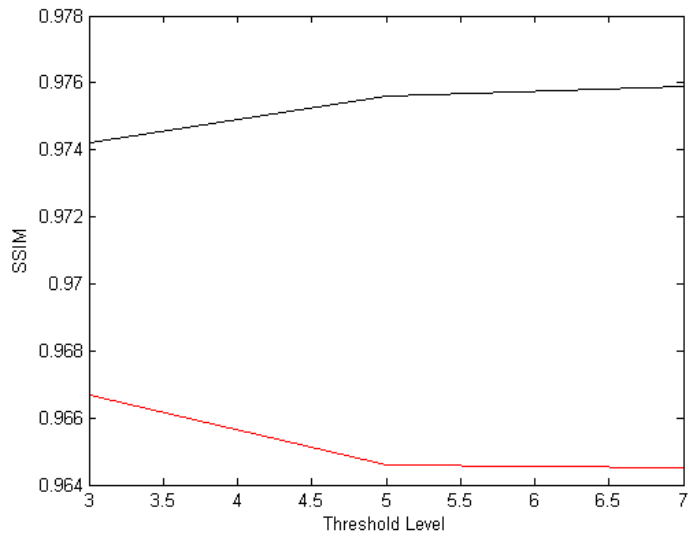

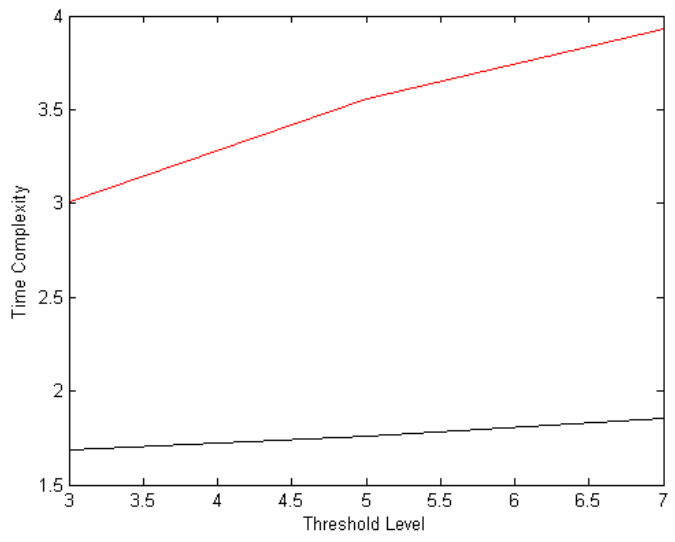

Fig.10. (j) Fig.10. (k) Fig.10. (l)

Fig.10. Results: (a,b,c) Comparison of PSNR, MSSIM and Time Complexity of House image between image segmentation in space domain and by proposed algorithm.

(d,e,f) Comparison of PSNR, MSSIM and Time Complexity of Lenna image between image segmentation in space domain and by proposed algorithm.

(g,h,i) Comparison of PSNR, MSSIM and Time Complexity of Pepper image between image segmentation in space domain and by proposed algorithm.

(j,k,l) Comparison of PSNR, MSSIM and Time Complexity of Baboon image between image segmentation in space domain and by proposed algorithm.

## 5. CONCLUSION

The performance of the proposed hybrid algorithm has been compared with the algorithm reported by Arora et al.[1]. Time taken by hybrid algorithm in wavelet domain is approximately half of the time taken by space domain algorithm. It can also be seen that segmentation done in wavelet domain gives improved PSNR compared to segmentation done by Arora et al. at same threshold level using mean and standard deviation. The number of thresholds required to reach the saturation PSNR is far less than thresholds required in space domain. SSIM of image segmented in wavelet domain is always better than the image segmented by using Arora et al. algorithm. Finally, more distinct regions can be observed in an image using wavelet domain segmentation compared to space domain segmentation. It is worth emphasizing that after the segmentation only 4 times the $n$ (number of thresholds) + 1 wavelet coefficients captures the significant variation of image.

## REFERENCES

[1] S. Arora, J. Acharya, A. Verma, P.K. Panigrahi, Multilevel thresholding for image Segmentation through a Fast Statistical Recursive Algorithm, Pattern Recognition Letters, 29 (2008) 119-125.

[2] R.C. Gonzales, Woods R.E. Digital Image Processing (2ed., PH, 2001)

[3] M. Sezgin, B. Sankur, Survey over image thresholding techniques and quantitative performance evaluation, Journal of Electronic Imaging, 13(1) (2004) 146-165.

[4] N.Otsu, A threshold selection using gray level histograms, IEEE Trans. Systems Man Cybernet. 9 (1979) 62-69

[5] J.N.Kapur, P.K Sahoo, A.K. Wong, A new method for gray-levelpicture thresholding using the entropy of the histogram, Computer Vision, Graphics and Image Processing, 29 (1985) 273-285.

[6] P.S. Liao, T.S. Chen, P.C. Chung, A fast Algorithm for Multilevel Thresholding. Journal of Information Science and Engineering, 2001, 713-727.

[7] A. S. Abutaleb, Automatic thresholding of gray level pictures using two dimensional entropy, Computer Vision Graphics Image Process. 47 (1989) 22-32.

[8] W. Niblack, An Introduction to Digital Image Processing, Prentice Hall, 1986 , 115-116.

[9] S. Hemachander, A. Verma, S. Arora, P.K. Panigrahi, Locally adaptive block thresholding method with continuity constraint, Pattern Recognition 28(2007), 119-124.

[10] S. S. Reddi, S. F. Rudin, H. R. Keshavan, An optical multiple threshold scheme for image segmentation, IEEE Trans. System Man and Cybernetics, 14 (1984) 661-665.

[11] T. W. Ridler, S. Calward, Picture Thresholding Using an Iterative Selection Method, IEEE Trans. Systems, Man and Cybernetics, 8 (1978) 630-632.

[12] C.C. Chang, L.L. Wang, A fast multilevel thresholding method based on lowpass and highpass filtering, Pattern Recognition Letters 18(1977), 1469-1478.

[13] Q. Huang, W. Gai, W. Cai, Thresholding technique with adaptive window selection for uneven lighting image. Pattern Recognition Letters 26(2005),801-808.

[14] S. Boukharouba, J.M. Rebordao, P.L. Wendel, An amplitude segmentation method based on the distribution function of an image. Comput. Vision Graphics Image Process. 29(1985), 47-59.

[15] J. Kittler, J. Illingworth, Minimum Error Thresholding, Pattern Recognition, 19(1986), 41-47.

[16] N. Papamarkos, B. Gatos, A new approach for multilevel threshold selection. Graphics Models Image Process. 56(1994), 357-370.

[17] K. Hammouche, M. Diaf, P. Siarry, A comparative study of various meta-heuristic techniques applied to the multilevel thresholding problem, Engineering Applications of Artificial Intelligence, 23 (2009), 676-688.

[18] Sachin P Nanavati and Prasanta K Panigrahi, 2005. Wavelets: Applications to Image Compression-I, Resonance, 10, 52-61.

[19] Sachin P Nanavati and Prasanta K Panigrahi, Wavelets: Applications to Image Compression-II, Resonance, 10(2005), 19-27.

[20] I. Daubechies, Ten Lectures on Wavelets, Vol. 61 of Proc. CBMS-NSF Regional Conference Series in Applied Mathematics, Philadelphia, PA: SIAM (1992).

[21] S.G. Mallat, A Wavelet Tour of Signal Processing. New York: Academic (1999).

[22] A Primer on wavelets and their scientific applications - James A Walker

[23] Zhou Wang, A.C. Bovik, H.R. Sheikh, E.P. Simoncelli, Image quality assessment: from error visibility to structural similarity, Center for Neural Sci., New York Univ., NY, USA, 13-4(2004), 600-612.